\documentclass[12pt]{article}
\usepackage{graphicx}
\graphicspath{{./figures/}{./}}
\usepackage[numbers]{natbib}
\usepackage{lineno} % used for \linenumbers
\usepackage{color}

\newcommand{\dder}[2]{\frac{\mathrm{d} #1}{\mathrm{d} #2}} % derivative
\newcommand{\D}{\mathrm{d}} % d in derivatives
\newcommand{\eq}[1]{(\ref{#1})} % equation references
\begin{document}

\def\spacingset#1{\renewcommand{\baselinestretch}%
{#1}\small\normalsize} \spacingset{1}

%%%%%%%%%%%%%%%%%%%%%%%%%%%%%%%%%%%%%%%%%%%%%%%%%%%%%%%%%%%%%%%%%%%%%%%%%%%%%%

\title{\bf Flow in Bounded and Unbounded Pore Networks with Different Connectivity}
\author{Daniel W.\ Meyer\thanks{The first author gratefully acknowledges Branko Bijeljic from the Imperial College London for providing network data. Moreover, he acknowledges financial support from ETH Z\"urich.}
\ and Artur Gomolinski\hspace{.2cm}\\
  Institute of Fluid Dynamics, ETH Z\"urich}
\maketitle

\bigskip
\begin{abstract}
This work is concerned with the intricate interplay between node or pore pressures and connection or throat conductivities in flow or pore networks. A setting similar to pore networks is given by fracture networks. Recently, a non-local generalization of Darcy's law for flow and transport in porous media was presented in the context of unbounded or periodic pore networks. In this work, we first outline a robust method for the extraction of the hydraulic conductivity distribution, which is at the heart of the non-local Darcy formulation. Second, a theory for mean pressure and flow in bounded networks is outlined. Predictions of that theory are validated against numerical network results and it is demonstrated that the theory works well for networks with high connectivity involving pores with high coordination numbers. For other networks, improvements to the outlined theory are proposed and their accuracy is assessed.
\end{abstract}

\noindent%
\unitlength\textwidth
\begin{figure}
\begin{picture}(1.0,0.01)
\put(0,1.4){\fbox{
\begin{minipage}[b]{0.99\textwidth}
The final form of this manuscript was published by Elsevier in the \textbf{Journal of Hydrology} on 27/07/2019 and is available online at https://doi.org/10.1016/j.jhydrol.2019.123937
\end{minipage}}}
\end{picture}
\end{figure}
\vfill

\newpage
%\spacingset{1.45} % DON'T change the spacing!
%\linenumbers

\section{Introduction}

\begin{figure}
\begin{center}
\includegraphics[width=0.8\textwidth]{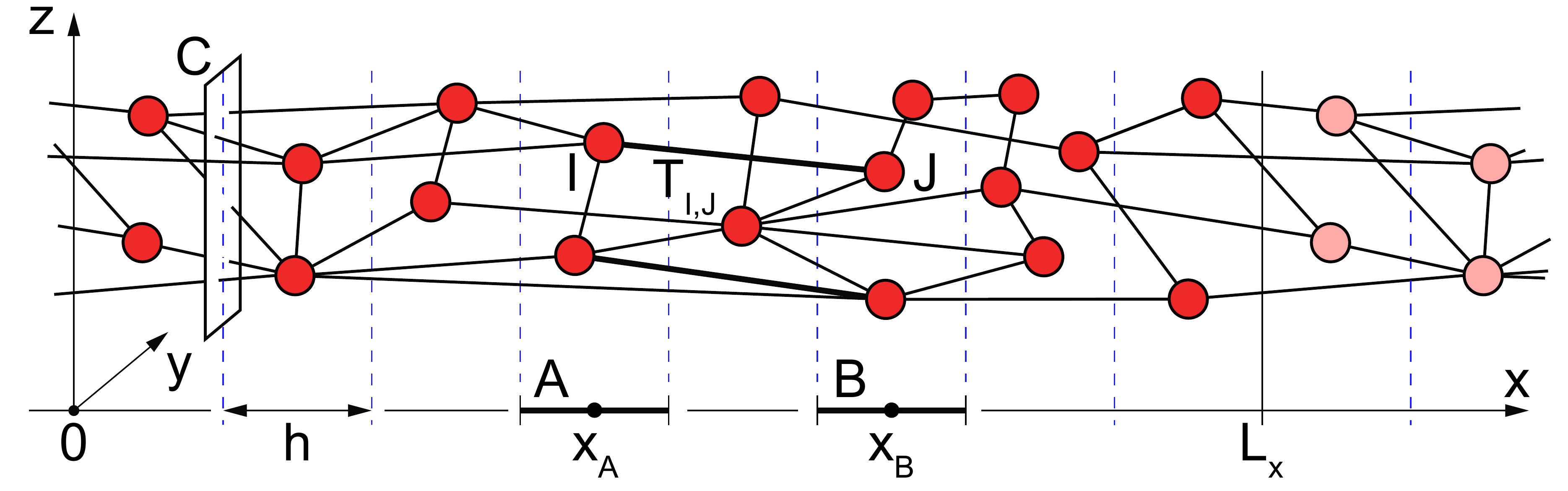}
\caption{Sketch of a pore network in $x$-$y$-$z$-space. Pores are depicted as red spheres and connecting throats as black lines. The cross-sectional area of the network in $y$-$z$-directions is represented by plane~$C$. Two slabs $A$ and $B$ perpendicular to the $x$-direction are centered at $x_A$ and $x_B$, respectively, and have a thickness~$h$. The network is $L_x$-periodic in $x$-direction with periodic pore copies depicted as light-red spheres.\label{figNetwork}}
\end{center}
\end{figure}%
Based on a porous-medium representation that idealizes the pore space in the form of a set of pores connected through a set of throats \citep[e.g.,][]{Constantinides:1989a,Blunt:1990a}, we have developed the non-local Darcy formulation \citep{Delgoshaie:2015a}. More recently, \citet{Jenny:2017a} have related this non-local formulation to a particle-based transport description, which generalizes existing continuous time random walk (CTRW) models. Unlike Darcy's law, where the flow at one point is determined by the local pressure gradient, the non-local formulation accounts for short- as well as long-range pressure transmission due to throats connecting pores over a range of distances. More specifically, for three-dimensional flow-networks that are statistically homogeneous in the spatial directions normal to the mean flow in $x$-direction (see figure~\ref{figNetwork}), the following expression for the averaged pressure $p(x)$ was derived
\begin{equation}\label{eqnlD}
p(x)\int_D T(x,x^\prime)dx^\prime - \int_D T(x,x^\prime)p(x^\prime) dx^\prime = q(x)
\end{equation}
\citep[see equation~(5) in][]{Delgoshaie:2015a}. In expression~\eq{eqnlD}, $p(x)$ is the average pore pressure with unit $[F/L^2]$ resulting from pores in a thin slab of thickness~$h$ centered at $x$.\footnote{Slabs are considered as regions of statistical space-stationarity with respect to flow and pressure.} More specifically based on figure~\ref{figNetwork}, $p(x_A) \equiv n_A^{-1} \sum_{I\in A} p_I$, where $n_A$ is the number of pores in slab~$A$, $p_I$ is the pressure in pore~$I$, and index~$I$ enumerates the subset of pores contained in slab~$A$. Similarly the specific source term $q(x)$ $[(L^3/T)/L^3 = 1/T]$ is defined as $q(x_A) \equiv (C h)^{-1} \sum_{I\in A} Q_I$ with $Q_I$ $[L^3/T]$ being the net-flow out of pore~$I$ ($Q_I > 0$) and $C$ being the cross-sectional area of the pore network perpendicular to~$x$. The conductivity distribution $T(x,x^\prime)$ $[L/(F T) = T/M]$ is defined as
\begin{equation}\label{eqT}
T(x_A,x_B) \equiv \frac{1}{C h^2}\frac{\sum_{I\in A}\sum_{J\in B} T_{I,J}(p_I - p_J)}{p(x_A) - p(x_B)},
\end{equation}
where $I$ and $J$ are indices of pores that reside in network slabs~$A$ and $B$, respectively, both having a thickness $h$ (see figure~\ref{figNetwork}). $T_{I,J}$ $[L^5/(F T)]$ is the hydraulic conductivity of the throat that connects pores~$I$ and $J$.\footnote{Since $T_{I,J} = T_{J,I}$, it follows that $T(x_A,x_B) = T(x_B,x_A)$.} (In the case of cylindrical throats, the throat conductivity $T_{I,J}$ is given as
\begin{equation}\label{eqTij}
T_{I,J} \equiv \frac{\pi D_{I,J}^4}{128 \mu L_{I,J}},
\end{equation}
where $D_{I,J}$ and $L_{I,J}$ are the diameter and length, respectively, of the throat connecting pores~$I$ and $J$. $\mu$ $[M/(LT)]$ is the dynamic viscosity of the fluid in the network.) In the non-local Darcy formulation~\eq{eqnlD}, $T(x,x^\prime)$ represents the non-local pressure transmission effect stated earlier. In the definitions of $p(x)$, $q(x)$, and $T(x,x^\prime)$, the slab thickness $h$ is assumed to be sufficiently small, such that variations in these quantities are resolved and at the same time large enough such that statistical errors inherent to finite pore counts remain small.

The conductivity distribution $T(x,x^\prime)$ directly reflects the connectivity of a porous medium, which is an aspect that has been investigated in the past by different groups. For example, \citet{Tyukhova:2015a} outline a methodology to identify a network of channels of least resistance or preferential flow in a Darcy-flow context. These channels result from an iterative deformation process that is driven by local differences in the heterogeneous hydraulic conductivity \citep[section~2.1]{Tyukhova:2015a}. In the followup contribution \citep{Tyukhova:2016a}, their methodology is combined with the multirate mass transfer model (MRMT) \citep{Haggerty:1995a} and model predictions are compared against reference simulations. Moreover, in the context of discrete fracture networks (DFN) \citep{Long:1982a,Robinson:1984a}, \citet{Maillot:2016a} compare commonly-used Poisson DFNs (comprised of statistically independent fractures) to kinematic fracture models (KFM), where the fracture pattern results from a growth process subject to fracture interaction. They find that the Poisson and KFM DFNs display quite different fracture connectivities leading to significant differences in flow and permeability.

Coming back to the non-local Darcy formulation, we could show in the limit of scales of interest being much larger than the longest throats in the network, that the formulation reduces to classical Darcy flow as
\begin{equation}\label{eqDarcyLimit}
-\frac{k}{\mu}\dder{^2 p(x)}{x^2} = q(x) \mbox{ with the permeability } k \equiv \mu \int_0^{\infty} s^2 T(s) \D s
\end{equation}
$[L^2]$ and the dynamic viscosity $\mu$ $[M/(LT)]$ of the fluid in the network \citep[Section~2.3]{Delgoshaie:2015a}.

For practical reasons, our past and current research efforts have been focusing on space-stationary settings, that is setups where $T$ does not depend on the actual points~$x$ and $x^\prime$, but solely on their separation distance $s \equiv x - x^\prime$. This, however, does not only require the network topology to be space stationary, but the pressure statistics have to be spatially independent as well (see expression~\eq{eqT}).

In our earlier work \citep{Delgoshaie:2015a}, the conductivity distribution was extracted based on a network, which was periodic in mean-flow $x$-direction. A mean flow was induced in the network by introducing designated in-/outflow slabs, respectively (see \citep[figure~4]{Delgoshaie:2015a}). In the indicated slabs the pore pressures were set to a constant high/low value respectively, which induced a driving pressure gradient in the remaining regions of the network (see figure~\ref{figPio} in this work). The pressure field, however, is different in the in-/outflow slabs (where equal pore pressures are prescribed) compared to the other regions of the network (variable pressures). For this reason, $T(s)$ was estimated via expression~\eq{eqT} from slabs that were separated from the in-/outflow slabs by buffer zones. The primary function of these buffer zones was the relaxation of inhomogeneities in flow and pressure. In this work, we present a space-stationary network setup similar to the one briefly mentioned in our earlier work \citep{Jenny:2017a}. The stationary setup will be discussed in detail and assessed with a validation study, where we compare the network permeability resulting via expression~\eq{eqDarcyLimit} from $T(s)$ against its counterpart calculable from global network quantities such as size, net flow, and pressure drop.

In addition to the space-stationary setup intended for the extraction of $T(s)$, a theory for non-local flow in samples with finite extensions is developed. This development implies that the flux through a finite pore network is composed of two components: the first component represents short-circuit throats that directly connect the in- and outflow boundaries without any intermediate pores. The second flux component is comprised of throats that connect the two boundaries through intermediate pores in the network. Both fluxes are calculable as a function of the network thickness~$L$ via integral expressions from the conductivity distribution $T(s)$. We validate predictions of our theory against numerical results stemming from networks of different topology.

After having outlined the non-local Darcy formulation in this introductory section, we present and validate the methodology for the construction of space-stationary pore networks and flow solutions in section~\ref{secPeriodicNetwork}. The theory for bounded networks is presented and assessed in section~\ref{secBoundedNetwork}.

\section{Space-Stationary Pore Networks and Flow}\label{secPeriodicNetwork}

\subsection{Space-Stationary Pore Networks}\label{subsecNetwork}

In order to imitate an unbounded space-stationary pore network, we resort to networks that are periodic in all spatial directions. For example in figure~\ref{figNetwork}, a network with periodic throats in $x$-direction is depicted. To generate such networks, we use a slightly generalized version of the algorithm outlined by \citet[Figure~3.3]{Idowu:2009c}. This algorithm accounts for correlation between throat thickness and size of connected pores as proposed earlier by \citep[e.g.,][]{Constantinides:1989a}. More specifically, a periodic network is resulting from the following steps:
\begin{enumerate}
\item Given is a base network in a cube volume that may stem from a sample of a natural porous medium \citep[e.g.,][Appendix~I]{Sochi:2007a}. The base network is composed of pores that are connected through throats. Each pore has a position, radius, and list of connecting throats. Each throat is determined based on a radius and a pair of connecting pores. Moreover, a target volume $[0,L_x]\times [0,L_y]\times [0,L_z]$ for the periodic network to be generated is prescribed.\footnote{To avoid in steps~4 to 6 the formation of throats that connect a pore with itself, the side lengths $L_x$, $L_y$, and $L_z$ of the target volume need to be larger than the length $L_m$ of the longest throat in the base network.}
\item With the same pore density as in the base network, pores are randomly placed in the target volume following a spatially uniform distribution. The properties of the newly placed pores such as radius and coordination number, i.e., the number of throats connected with a certain pore, are assigned by randomly drawing (with replacement) pores from the base network.
\item To establish periodic throat connections reaching over boundaries of the target volume, pores located in the slices $0 \le x < L_m$ and $L_x-L_m \le x < L_x$ are copied/shifted in $x$-direction outside the target volume by $+L_x$ and $-L_x$, respectively (see figure~\ref{figNetwork}). Here, $L_m$ is the length of the longest throat in the base network. The copy/shift operation is repeated in $y$- and subsequently in $z$-direction while including previously copied pores.\label{enumBuffer}
\item The first pore in the target volume with index $I = 1$ and coordination number~$c_1$ is connected through throats to the $c_1$ closest neighboring pores. This step is repeated for the other pores with indices $I = 2,3,\ldots$ in the target volume while accounting for existing connections of neighboring pores. More specifically, a neighboring pore~$J$, which has already been connected to $c_J$ pores or pore copies, is not connected to pore~$I$. Instead the search neighborhood for connection candidates to pore~$I$ is expanded up to a maximal distance $L_m$ from pore~$I$.\label{enumConnectPores}
\item Throat connections involving a pore copy (resulting from step~\ref{enumBuffer}) are rerouted to the original pore while storing the throat length that resulted from the location of the pore copy. With all throats connecting now pores in the target volume, the pore copies from step~\ref{enumBuffer} are removed.
\item Throat radii are assigned such that throats connecting pores with large radii have a large radius as well. More specifically, for each throat connection that resulted from step~\ref{enumConnectPores} the radii of the connected pores are summed and a throat is randomly drawn from the base network. The radius of the throat with the largest pore-radius sum is set to the largest throat radius from all randomly drawn throats. Similarly, the radius of the throat with the second-largest pore-radius sum is set to the second-largest throat radius from the base network and so on for the remaining throat radii.
\end{enumerate}

\subsection{Space-Stationary Flow}\label{subsecFlow}

Given the periodic network topology resulting from the algorithm outlined in the previous section, a space-stationary flow is induced based on a decomposition of the pressure into a sloping mean field and a periodic pressure fluctuation. (The same idea has been successfully applied in the context of Darcy flow in formations with heterogeneous permeability distributions \citep[e.g.,][]{Cirpka:2002a,Meyer:2018b}.) The corresponding development is based on the flux balance of an arbitrary pore~$I$:
\begin{equation}\label{eqQ}
\sum_{J} Q_{I,J} = 0 \ \forall\ I \mbox{ with } Q_{I,J} \equiv T_{I,J} (p_I - p_J)
\end{equation}
and where index~$J$ enumerates the pores that are connected through a throat with conductivity $T_{I,J}$ to pore~$I$. The pore pressure $p_I$ is now decomposed into a fluctuating part $p_I^\prime$ and a mean part $\langle p(x_I)\rangle$, i.e., $p_I = p_I^\prime + \langle p(x_I)\rangle$, where the latter is linearly varying in $x$-direction, i.e.,
\begin{equation}\label{eqpm}
\langle p(x)\rangle = P\left(1-\frac{x}{L_x}\right),
\end{equation}
and determined by the pressure magnitude~$P$. Insertion of this decomposition into flux balance~\eq{eqQ} leads to
\begin{equation}\label{eqps}
\sum_{J} T_{I,J} (p_I^\prime - p_J^\prime) = \frac{P}{L_x} \sum_{J} T_{I,J} (x_I - x_J) \ \forall\ I,
\end{equation}
which is an equation system for the pressure fluctuations~$p_I^\prime$. In order to fix the pressure level of the fluctuations, equation system~\eq{eqps} has to be modified for example by replacing the flux balance of pore $I = 1$ with the auxiliary condition $p_1^\prime = 0$. Note that the throat fluxes $Q_{I,J}$ are determined based on total pore pressures and therefore are proportional to~$P$.

\subsection{Extraction of Conductivity Distribution}\label{subsecT}

\subsubsection{Formulation}

From the network and flow topology resulting from the techniques outlined in the previous two sections, the conductivity distribution~\eq{eqT} can be extracted by decomposing the network into slabs of thickness~$h$ as sketched in figure~\ref{figNetwork}. However, since both the network and the flow topology are spatially independent, for example $T(s \equiv x_A-x_B) = T(x_A,x_B)$ can be estimated not only from throats connecting two particular slabs centered at $x_A$ and $x_B$, but from all throats connecting $L_x/h$ pairs of slabs separated by $s \equiv x_A-x_B$. A corresponding adaptation of expression~\eq{eqT} leads to
\begin{equation}\label{eqTs}
T(s) = \frac{1}{C h^2}\frac{h}{L_x}\frac{\displaystyle \sum_{i = 1}^{L_x/h} \; \sum_{i-1 \le x_I/h < i} \;\; \sum_{i-1 \le (x_J-s)/h < i} T_{I,J}(p_I - p_J)}{s\,P/L_x},
\end{equation}
with separations~$s$ being even multiples of $h$. In expression~\eq{eqTs}, the slab pressure difference from the denominator of equation~\eq{eqT} was determined based on the prescribed mean pressure drop $P/L_x$. Since the pore pressures $p_I$ and $p_J$ in equation~\eq{eqTs} are proportional to the magnitude~$P$ (see expressions~\eq{eqpm} and \eq{eqps}), the effect of $P$ on $T(s)$ cancels.

\subsubsection{Convergence}\label{subsubsecConv}

\begin{figure}
\begin{center}
\includegraphics[width=0.6\textwidth]{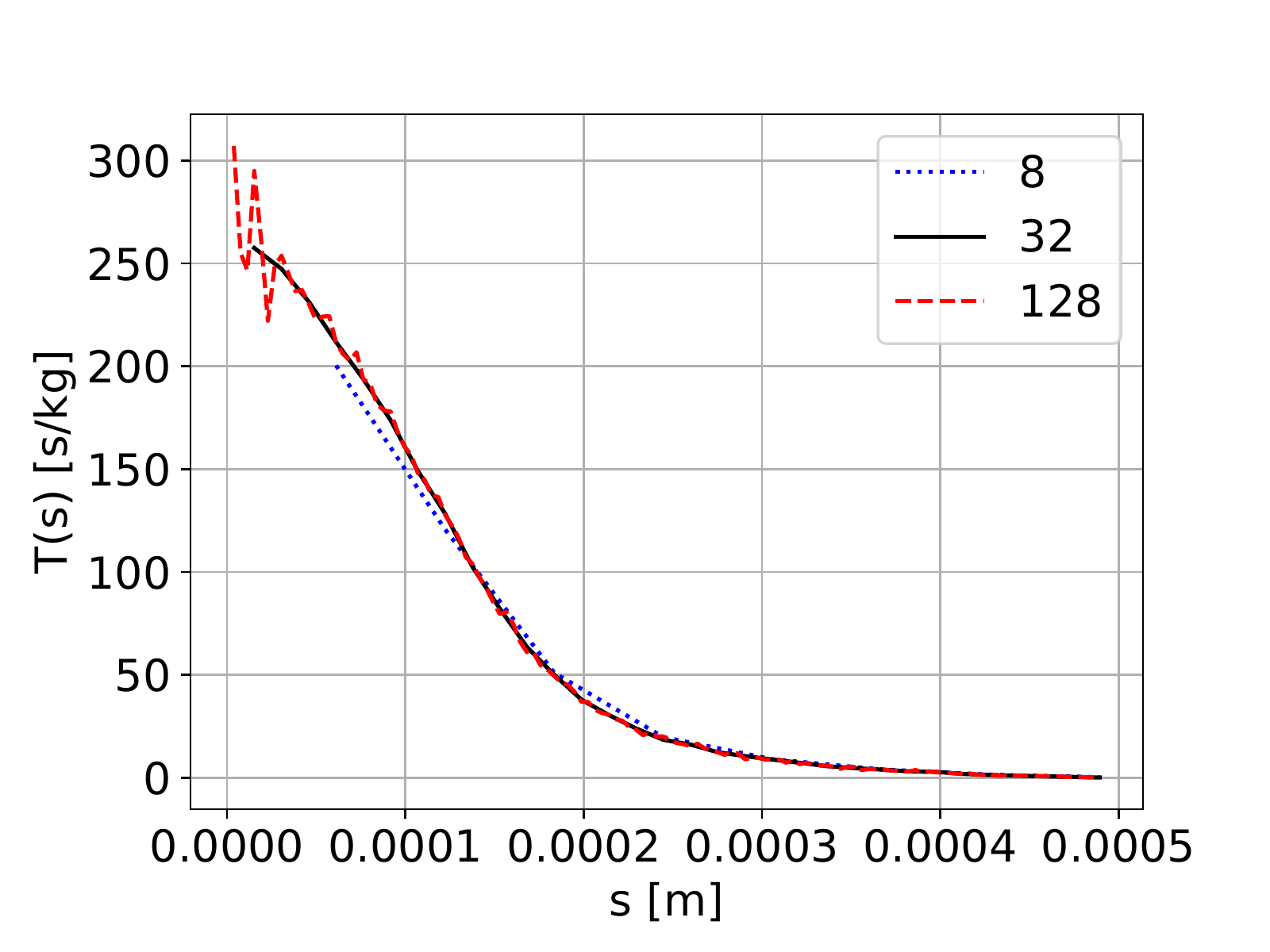}
\caption{Conductivity distributions $T(s)$ resulting from a sandstone network of size $(14\mbox{mm})^3$. Distributions resulting from different slab thicknesses~$h$ are depicted, i.e., $h = L_m/n$ with $L_m = 4.89\times 10^{-4}\mbox{m}$ and $n = 8$, 32, and 128.\label{figTcube}}
\end{center}
\end{figure}%
Based on the sandstone pore-network data-set available in \citep{Idowu:2009b} and the algorithm from section~\ref{subsecNetwork}, a periodic pore network of dimensions $(14\mbox{mm})^3$ with roughly 1.36 million pores and 2.9 million throats was generated. Next, a constant mean flow was induced using the setup outlined in section~\ref{subsecFlow} with a dynamic viscosity $\mu = 8.9\times 10^{-4} \mbox{kg/(ms)}$ and conductivity distributions were extracted by using expression~\eq{eqTs} with different slab thicknesses $h = L_m/n$. The resulting distributions for $n = 8$, 32, and 128 are depicted in figure~\ref{figTcube}. While $n = 8$ slabs resolve the variability in $T(s)$ only poorly, 128 slabs lead to noisy estimates for $T(s)$ given on average $1.36\times 10^6 h/L_x \approx 373$ pores per slab. With 32 slabs per maximal throat length~$L_m$, the conductivity distribution is well resolved and statistical noise is kept at a low level. Based on that optimal $T(s)$ and expression~\eq{eqDarcyLimit} numerically approximated with a trapezoidal rule, the permeability $k = 3.86\times 10^{-13}\mbox{m}^2$ is calculable. The same permeability is obtained from Darcy's law with the total volumetric flux in $x$-direction $Q_x$ and the pressure gradient $P/L_x$ via
\begin{equation}\label{eqk}
k = \mu \frac{Q_x}{L_y L_z}\frac{L_x}{P}.
\end{equation}
This exact agreement confirms the Darcy limit~\eq{eqDarcyLimit} implied by the non-local Darcy theory.

\subsubsection{Comparison with In-/Outflow Slab Method}

\begin{figure}
\begin{center}
\includegraphics[width=0.6\textwidth]{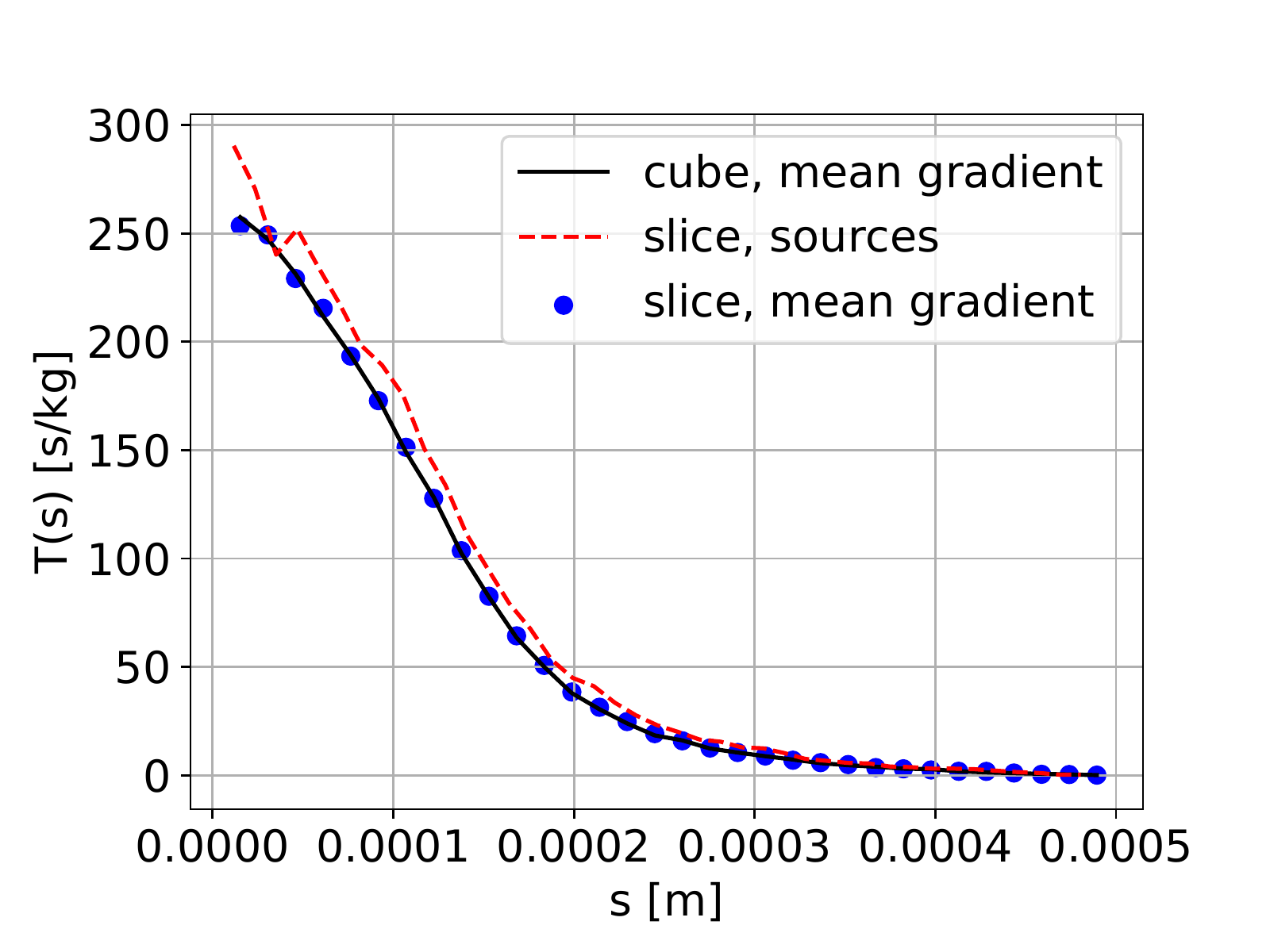}
\caption{Conductivity distributions $T(s)$ resulting from sandstone networks of sizes $(14\mbox{mm})^3$ (black solid) and $3\mbox{mm}\times 30\mbox{mm}\times 30\mbox{mm}$ (red dashed, blue dots) with throats of maximal length $L_m = 4.89\times 10^{-4}\mbox{m}$. The distributions (black solid, blue dots) resulted from the mean-pressure-gradient setup outlined in sections~\ref{subsecNetwork} and~\ref{subsecFlow}, while (red dashed) is based on the in-/outflow slab setup proposed earlier \citep{Delgoshaie:2015a}.\label{figTslice}}
\end{center}
\end{figure}
\begin{figure}
\begin{center}
\includegraphics[width=\textwidth]{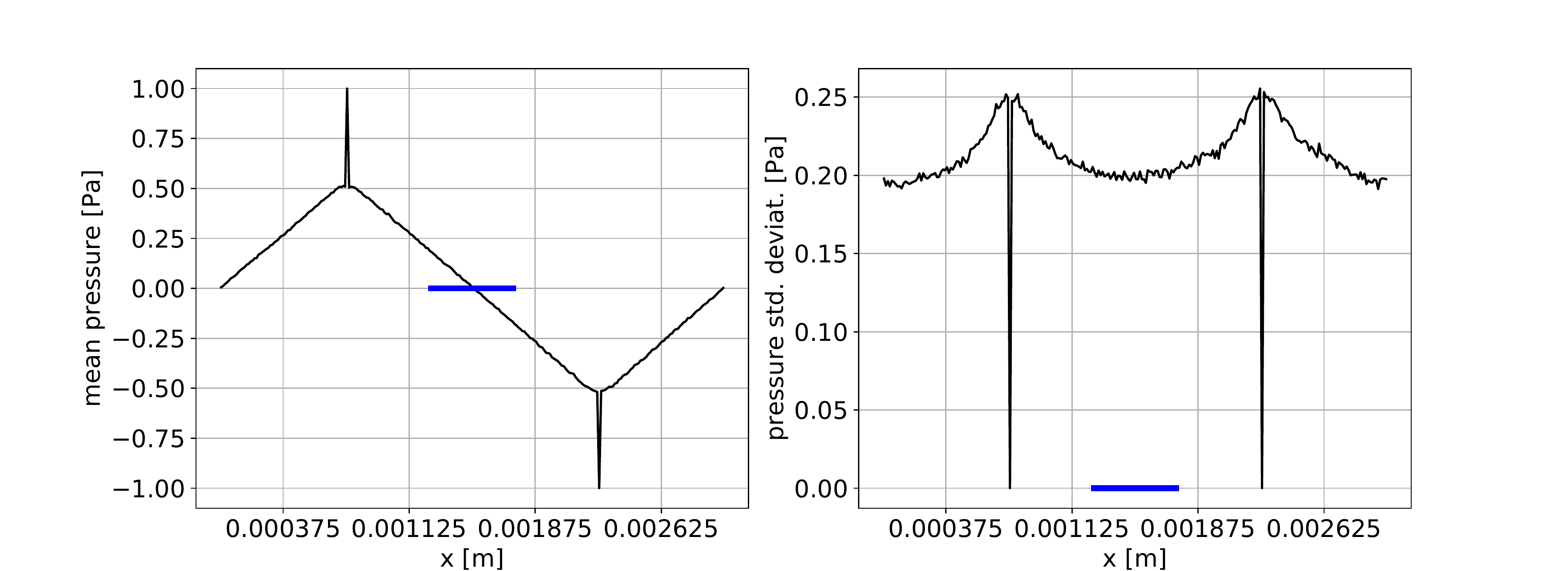}
\caption{Pore pressure statistics as a function of the mean-flow-parallel $x$-direction in a sandstone network of extensions $3\mbox{mm}\times 30\mbox{mm}\times 30\mbox{mm}$ with the in-/outflow slab setup. In the left panel, mean pore pressures averaged over 256 equidistantly-spaced slabs are plotted. In the right panel, corresponding standard deviations are provided. The gray vertical grid lines represent the bounds of the buffer layers placed around the in-/outflow slabs with high/low constant pore pressures, respectively. For scale, the thick blue line corresponds to the length~$L_m$ of the longest throat in the network.\label{figPio}}
\end{center}
\end{figure}

%
In a next step, we compare the previously computed conductivity distribution against the one resulting from our earlier setup with in-/outflow slabs \citep{Delgoshaie:2015a}, which induces a non-stationary flow field (see figure~\ref{figPio}). This comparison will reveal two inconsistencies. Like in our earlier contribution \citep{Delgoshaie:2015a}, a pore network slice of extensions $3\mbox{mm}\times 30\mbox{mm}\times 30\mbox{mm}$ was generated using the previously outlined algorithm and the sandstone dataset from \citet{Idowu:2009b}. The resulting network is comprised of 1.34 million pores and 2.85 million throats.\footnote{The maximum throat length $L_m$ remains unchanged. Furthermore, $L_x/L_m \approx 6$.} Firstly, we apply a mean pressure gradient in line with section~\ref{subsecFlow} and determine $T(s)$ based on expression~\eq{eqTs}. The resulting distribution is depicted in figure~\ref{figTslice} (blue dots) and is, despite the different network extensions, in agreement with the one from the cubical network discussed in the previous section~\ref{subsubsecConv} (solid black line).

For the estimation of the conductivity distribution with the in-/outflow slab method, the in- and outflow slabs were embedded in buffer layers of width $L_x/4 \approx 1.5L_m$ (see left panel in figure~\ref{figPio}). Therefore, when estimating $T(s)$ only throats located outside of these layers in the remaining half of the network volume are accounted for. The resulting $T(s)$ is included in figure~\ref{figTslice} through the red dashed line. It is seen that despite the buffer layers, inhomogeneities originating from the in-/outflow slabs induce a bias in the conductivity distribution $T(s)$ when comparing against the space-stationary setup (blue dots). As a result, the permeability $k = 4.64\times 10^{-13}\mbox{m}^2$, that results via expression~\eq{eqDarcyLimit} from $T(s)$, deviates from $k = 3.86\times 10^{-13}\mbox{m}^2$ reported in connection with the space-stationary setup. Moreover, $k = 4.64\times 10^{-13}\mbox{m}^2$ is inconsistent with $k = 4.19\times 10^{-13}\mbox{m}^2$, where the latter was calculated with equation~\eq{eqk} based on the net flux $Q_x$ in $x$-direction and with $P$ representing the pressure magnitude applied at the in-/outflow pores with positive/negative sign, respectively.

\begin{figure}
\begin{center}
\includegraphics[width=\textwidth]{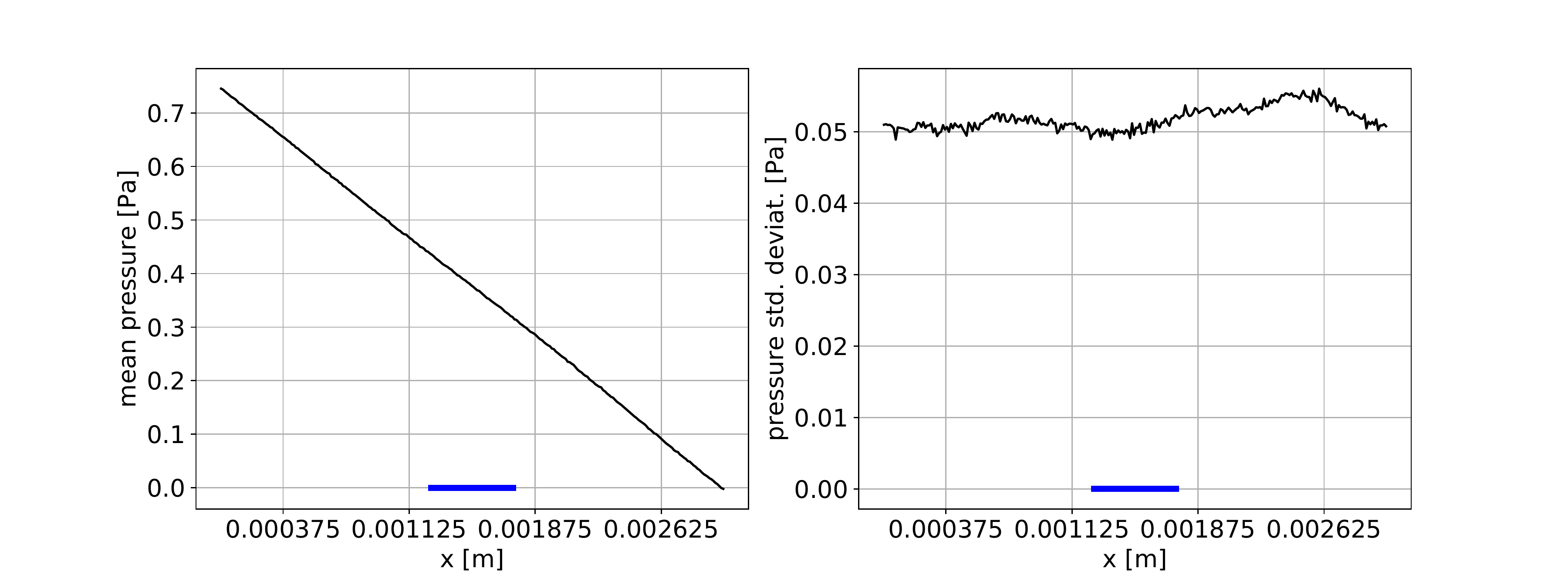}
\caption{Pore pressure statistics as a function of the mean-flow-parallel $x$-direction in a sandstone network of extensions $3\mbox{mm}\times 30\mbox{mm}\times 30\mbox{mm}$ with the space-stationary flow setup. In the left panel, mean pore pressures averaged over 256 equidistantly spaced slabs are plotted. In the right panel, corresponding standard deviations are provided. For scale, the thick blue line corresponds to the length~$L_m$ of the longest throat in the network.\label{figPs}}
\end{center}
\end{figure}%
To further illustrate the difference between the space-stationary and the in-/outflow slab setups, pressure statistics are reported in figures~\ref{figPio} and~\ref{figPs}, respectively. In the in-/outflow slabs located at $x = L_x/4$ and $3L_x/4$, respectively, constant pore pressures with zero variance were prescribed, which induce approximately linear pressure drops in large parts of the network, but at the same time inhomogeneities in the pore pressure statistics as shown in figure~\ref{figPio}. The newly advocated setup, on the other hand, displays a constant mean pressure drop in the whole network (figure~\ref{figPs} left panel) and periodic pore-pressure fluctuations with no systematic dependence on the downstream coordinate~$x$ (right panel). For networks of increasing size~$L_x$ and buffer layer thickness $L_x/4$, it is expected that the flow field inhomogeneities in regions outside of the buffer layers will vanish and $T(s)$ from the in-/outflow method will converge to the distribution from the space-stationary setup.

\section{Bounded Pore Networks}\label{secBoundedNetwork}

\begin{figure}
\begin{center}
\includegraphics[width=0.7\textwidth]{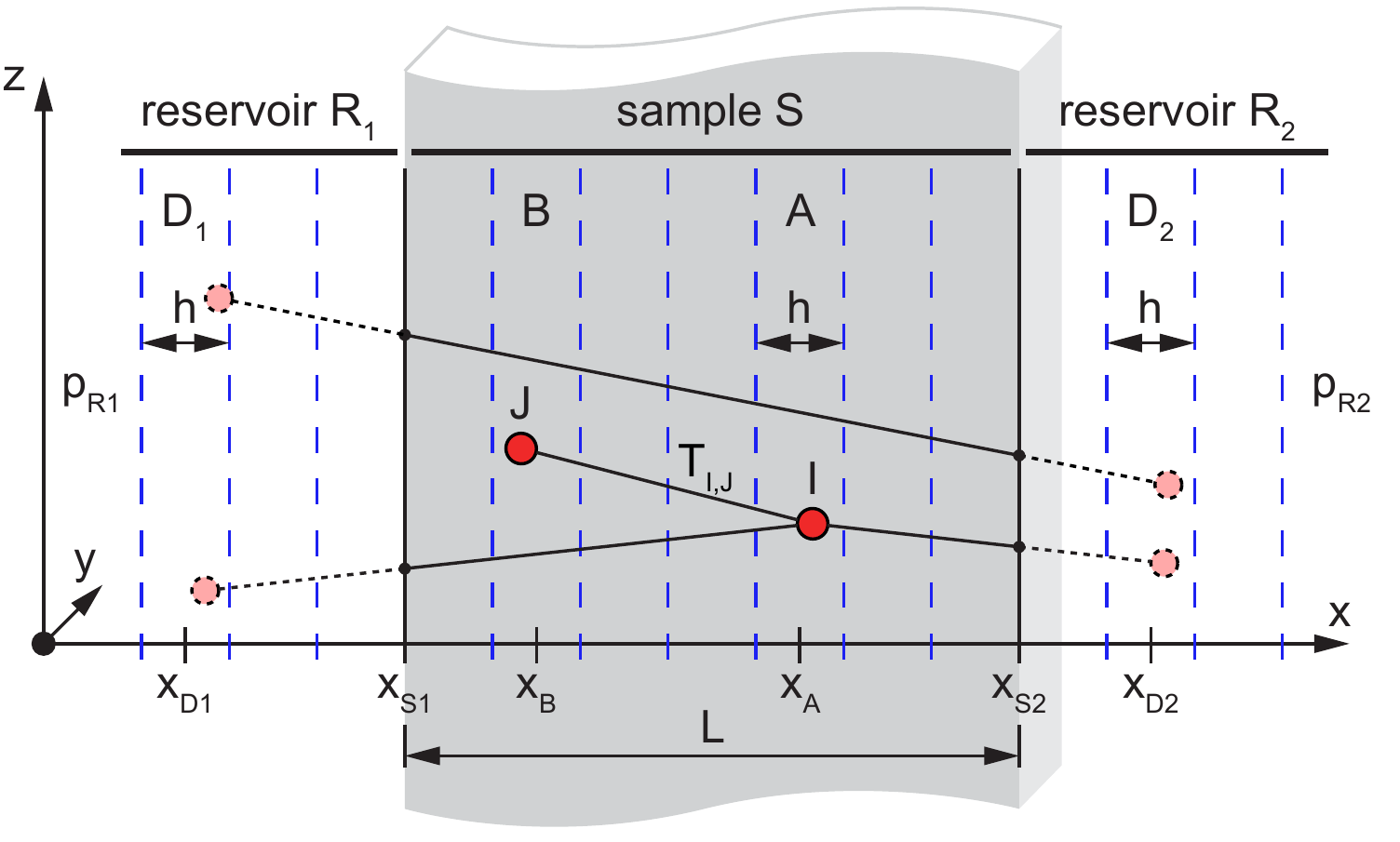}
\caption{Sketch of a cut pore network in $x$-$y$-$z$-space. The network is cut at positions $x_{S1}$ and $x_{S2}$ leading to a network sample~$S$ of thickness~$L$ (gray shaded block). The sample~$S$ is supplied by a liquid at the interfaces $x_{S1}$ and $x_{S2}$ through reservoirs~$R_1$ and $R_2$, respectively (with pressures $p_{R1}$ and $p_{R2}$). Exemplary pores are depicted as red spheres and connecting throats as black solid lines. Throats and pores that were cut away from the original network are shown with dotted lines. Slabs~$A$, $B$, $D_1$, and $D_2$ perpendicular to the $x$-direction and of thickness~$h$ are depicted using blue dashed lines.\label{figBoundedNetwork}}
\end{center}
\end{figure}%
After having outlined a method for the unbiased estimation of the conductivity distribution in space-stationary settings, we turn in this section to a theoretical extension of the formulation for bounded networks. Based on the terminology from figure~\ref{figBoundedNetwork}, we decompose the net flux $[L^3/T]$ from slab~$A$ into a component going to other slabs in network sample~$S$, and two components going to reservoirs~$R_1$ and $R_2$ (pores cut away):
\begin{eqnarray}\label{eqQA}
\lefteqn{Q_A = \sum_{I\in A} \sum_{J\in R_1\cup S\cup R_2} T_{I,J}(p_I - p_J)} \\
 & & = \sum_{I\in A}\left[\sum_{B\in S}\sum_{J\in B} T_{I,J}(p_I - p_J) + \sum_{D_1\in R_1}\sum_{J\in D_1} \frac{x_J-x_I}{x_{S1}-x_I} T_{I,J}(p_I - p_{R1}) \right. \nonumber \\
 & & \left. \hspace{12.4em} + \sum_{D_2\in R_2}\sum_{J\in D_2} \frac{x_J-x_I}{x_{S2}-x_I}T_{I,J}(p_I - p_{R2})\right] \nonumber \\
 & & = \sum_{B\in S}\sum_{I\in A}\sum_{J\in B} T_{I,J}(p_I - p_J) + \sum_{D_1\in R_1}\sum_{I\in A}\sum_{J\in D_1} \underbrace{\frac{x_J-x_I}{x_{S1}-x_I}}_{\displaystyle\hspace{-1ex}\approx\frac{x_{D1}-x_A}{x_{S1}-x_A}\hspace{-1ex}} T_{I,J}(\underbrace{p_I}_{\displaystyle\hspace{-1em}\approx p(x_A)\hspace{-1em}} - p_{R1}) + \ldots \nonumber \\
 & & = Ch \sum_{B\in S} T(x_A,x_B)[p(x_A)-p(x_B)]h + \sum_{D_1\in R_1} \frac{x_{D1}-x_A}{x_{S1}-x_A} \underbrace{\sum_{I\in A}\sum_{J\in D_1} T_{I,J}}_{\displaystyle\hspace{-1em}\approx Ch^2 T(x_A,x_{D1})\hspace{-1em}}[p(x_A) - p_{R1}] + \ldots \nonumber \\
 & & = C\left[\int_{x_{S1}}^{x_{S2}} T(x_A,x)[p(x_A)-p(x)] \D x + [p(x_A) - p_{R1}] \int_{-\infty}^{x_{S1}} \frac{x-x_A}{x_{S1}-x_A}T(x_A,x) \D x\right. \nonumber \\
 & & \left. \hspace{15.9em} +  [p(x_A) - p_{R2}] \int_{x_{S2}}^\infty \frac{x-x_A}{x_{S2}-x_A}T(x_A,x) \D x\right] h = 0. \nonumber
\end{eqnarray}
In the first step of the derivation of flux~$Q_A$, the domain $R_1\cup S \cup R_2$ was decomposed into slabs residing in sample~$S$ and reservoirs~$R_1$ and $R_2$. At the same time the conductivities $T_{I,J}$ of cut throats were corrected by their reductions in length consistent with expression~\eq{eqTij} for cylindrical throats. In the remaining steps, the definition of the conductivity distribution~\eq{eqT} was used and it was assumed that the pore pressures $p_I$ and $p_J$ can be approximated with their respective slab mean pressures, e.g., $p_I\approx p(x_A)$. (The validity of this assumption is assessed in section~\ref{subsecNetworks}.) In the last step of derivation~\eq{eqQA}, the Riemann sums over slabs in regions~$S$, $R_1$, and $R_2$ were replaced for slab thicknesses $h \to 0$ by integrals \citep[equation~(4)]{Delgoshaie:2015a}. Summation of all fluxes~$Q_A$ over all slabs~$A$ in sample~$S$ leads to
\begin{eqnarray}\label{eqQS}
\sum_{A\in S} \frac{Q_A}{C} = \hspace{12em} & & \\
\int\!\!\!\int_{x_{S1}}^{x_{S2}} T(x^\prime,x)[p(x^\prime)-p(x)] \D x\D x^\prime & \!\!\!+\!\!\! & \int_{x_{S1}}^{x_{S2}} [p(x^\prime) - p_{R1}] \int_{-\infty}^{x_{S1}} \frac{x-x^\prime}{x_{S1}-x^\prime}T(x^\prime,x) \D x \D x^\prime \nonumber \\
 & \!\!\!+\!\!\! & \int_{x_{S1}}^{x_{S2}} [p(x^\prime) - p_{R2}] \int_{x_{S2}}^\infty \frac{x-x^\prime}{x_{S2}-x^\prime}T(x^\prime,x) \D x \D x^\prime = 0. \nonumber
\end{eqnarray}
With $T(x,x^\prime) = T(x^\prime,x)$, the first two-dimensional integral on the right hand side vanishes, since contributions from point pairs $(x_1,x_2)$ and $(x_2,x_1)$ in the quadratic $x$-$x^\prime$ integration region cancel as $T(x_1,x_2)[p(x_1)-p(x_2)] = -T(x_1,x_2)[p(x_2)-p(x_1)]$. The remaining two integrals represent the net fluxes from the sample~$S$ into the reservoirs at pressures $p_{R1}$ and $p_{R2}$, respectively.

\subsection{Mean Pore-Pressure Drop in the Sample}\label{subsecPm}

In a next step, we verify whether the linear mean pressure drop
\begin{equation}\label{eqpx}
p(x) = p_{R1} + \frac{x-x_{S1}}{L}(p_{R2}-p_{R1})
\end{equation}
is a solution of equation~\eq{eqQS}. Insertion into equation~\eq{eqQS} leads to
\begin{eqnarray*}
\int_{x_{S1}}^{x_{S2}} (x^\prime - x_{S1}) \int_{-\infty}^{x_{S1}} \frac{x-x^\prime}{x_{S1}-x^\prime}T(x^\prime,x) \D x \D x^\prime & \!\!\!+\!\!\! & \int_{x_{S1}}^{x_{S2}} (x^\prime - x_{S2}) \int_{x_{S2}}^\infty \frac{x-x^\prime}{x_{S2}-x^\prime}T(x^\prime,x) \D x \D x^\prime \\
 = \int_{x_{S1}}^{x_{S2}} \left[\int_{-\infty}^{x_{S1}} (x-x^\prime)T(x^\prime,x) \D x \right. & \!\!\!+\!\!\! & \left. \int_{x_{S2}}^\infty (x-x^\prime)T(x^\prime,x) \D x \right] \D x^\prime \\
 = \int_{x_{S1}}^{x_{S2}} \left[\int_{-\infty}^\infty (x-x^\prime)T(x^\prime,x) \D x \right. & \!\!\!-\!\!\! & \left. \int_{x_{S1}}^{x_{S2}} (x-x^\prime)T(x^\prime,x) \D x \right] \D x^\prime,
\end{eqnarray*}
where the first inner integral is equal to zero since it can be written as $\int_{-\infty}^\infty x^{\prime\prime} T(x^\prime,x^\prime+x^{\prime\prime}) \D x^{\prime\prime} = \int_{-\infty}^\infty x^{\prime\prime} T(|x^{\prime\prime}|) \D x^{\prime\prime}$ for a space-stationary network with $T(x,x^\prime) = T(|x-x^\prime|)$. Moreover, the second integral is equal to zero based on an analogous argument made after equation~\eq{eqQS} and thus the linear pressure drop~\eq{eqpx} indeed satisfies equation~\eq{eqQS} under the condition of a space-stationary conductivity distribution.

\subsection{Flux from Sample to Reservoir}\label{subsecQSR2}

\begin{figure}
\begin{center}
\includegraphics[width=0.7\textwidth]{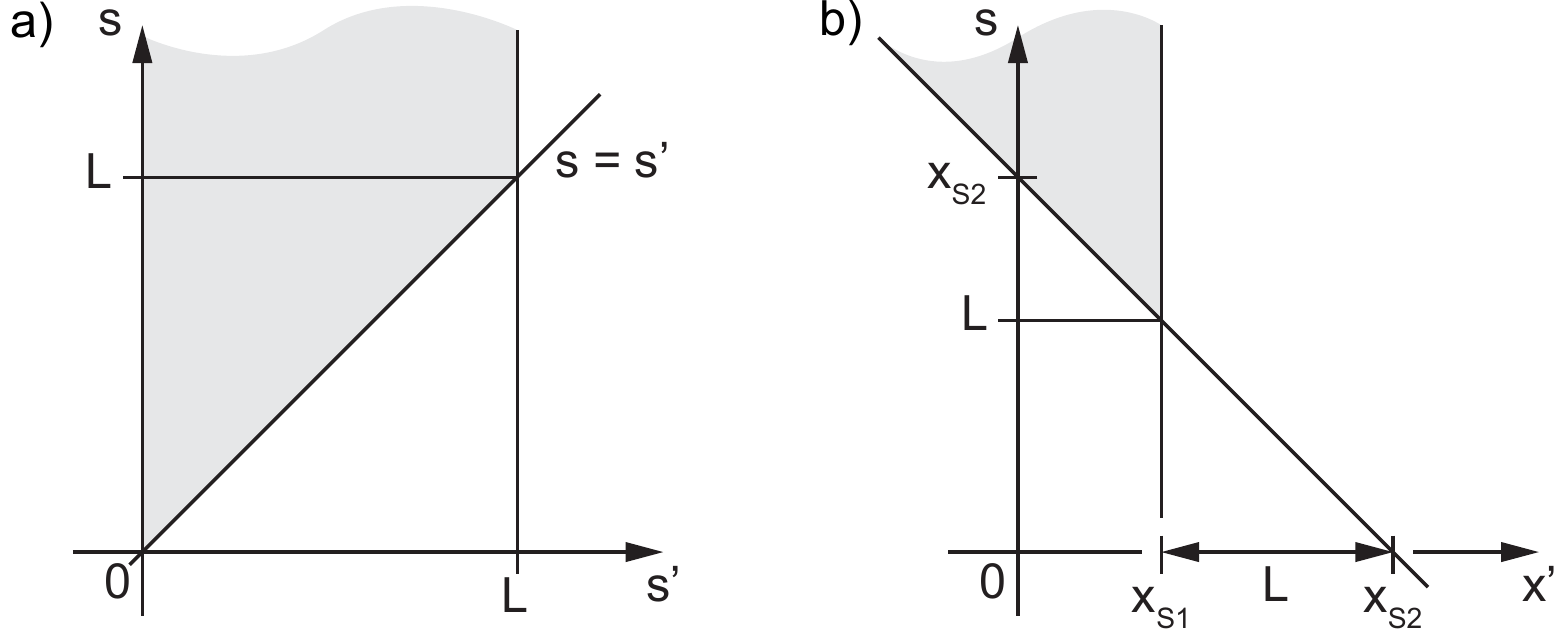}
\caption{Gray-shaded integration regions in $s^\prime$-$s$ and $x^\prime$-$s$ coordinate systems.\label{figIntRegion}}
\end{center}
\end{figure}%
Based on the mean pressure drop found in the previous section, a compact expression for the flux per unit area $q_{S,R1}$ $[L^3/(TL^2)]$ going from sample~$S$ to, e.g., reservoir~$R_1$ shall be derived: Starting point is the corresponding integral identified in equation~\eq{eqQS} and written here with a space-stationary conductivity distribution,
\begin{eqnarray}
q_{S,R1} & = & \int_{x_{S1}}^{x_{S2}} [p(x^\prime) - p_{R1}] \int_{-\infty}^{x_{S1}} \frac{x-x^\prime}{x_{S1}-x^\prime}T(x-x^\prime) \D x \D x^\prime \nonumber \\
 & = & \int_{x_{S1}}^{x_{S2}} \frac{p(x^\prime) - p_{R1}}{x_{S1}-x^\prime} \int_{-\infty}^{x_{S1}-x^\prime} s\, T(s) \D s \D x^\prime \nonumber \\
 & = & \int_0^L \frac{p(s^\prime+x_{S1}) - p_{R1}}{-s^\prime} \int_{-\infty}^{-s^\prime} s\, T(s) \D s \D s^\prime \nonumber \\
 & = & \int_0^L [p(s^\prime+x_{S1}) - p_{R1}]\frac{1}{s^\prime} \int_{s^\prime}^\infty s\, T(s) \D s \D s^\prime.
\end{eqnarray}
With the linear mean pressure drop~\eq{eqpx} we obtain
\begin{equation}\label{eqqSR1}
q_{S,R1} = \frac{p_{R2}-p_{R1}}{L} \int_0^L \int_{s^\prime}^\infty s\, T(s) \D s \D s^\prime.
\end{equation}
After switching the integration order based on figure~\ref{figIntRegion}a, the double integral in expression~\eq{eqqSR1} can be rewritten as
\begin{eqnarray}\label{eqqSR12}
\int_0^L \int_{s^\prime}^\infty s\, T(s) \D s \D s^\prime & \!\!\!=\!\!\! & \int_0^\infty \int_0^{\min(s,L)} s\, T(s) \D s^\prime \D s = \int_0^\infty \min(s,L) s\, T(s) \D s \nonumber \\
 & \!\!\!=\!\!\! & \int_0^L s^2 T(s) \D s + \int_L^\infty Ls T(s) \D s,
\end{eqnarray}
which will be useful later.

\subsection{Short-Circuit Flux between Reservoirs}\label{subsecQR1R2}

So far, the derivations in section~\ref{secBoundedNetwork} have focused on flow through  sample~$S$ via pores located in the sample. However, there is a second component to the flow between reservoirs~$R_1$ and $R_2$ through sample~$S$. As illustrated in figure~\ref{figBoundedNetwork}, if $L < L_m$, there may be throats that go straight through sample~$S$, short-circuiting reservoirs~$R_1$ and $R_2$. Based on similar arguments as applied in the derivation of equation~\eq{eqQA}, the short-circuit flux is estimated in the following. The flux between two slabs~$D_1$ and $D_2$ is given as 
\begin{equation}\label{eqQD1D2}
Q_{D1,D2} = C \frac{x_{D2}-x_{D1}}{L} T(x_{D1}-x_{D2}) (p_{R1}-p_{R2}) h^2
\end{equation}
and summation of all pairs of slabs~$D_1$ and $D_2$ in reservoirs~$R_1$ and $R_2$, respectively, leads to the net flux between $R_1$ and $R_2$ bypassing pores in samples~$S$, i.e.,
\begin{eqnarray}\label{eqqR1R2}
q_{R1,R2} & = & \frac{1}{C}\sum_{D_1\in R_1}\sum_{D_2\in R_2} Q_{D1,D2} \nonumber \\
 & = & \frac{p_{R1}-p_{R2}}{L} \int_{-\infty}^{x_{S1}} \int_{x_{S2}}^\infty (x-x^\prime) T(x-x^\prime) \D x  \D x^\prime \nonumber \\
 & = & \frac{p_{R1}-p_{R2}}{L} \int_{-\infty}^{x_{S1}} \int_{x_{S2}-x^\prime}^\infty s T(s) \D s \D x^\prime \nonumber \\
 & = & \frac{p_{R1}-p_{R2}}{L} \int_L^{\infty} \int_{x_{S2}-s}^{x_{S1}} s T(s) \D x^\prime \D s \nonumber \\
 & = & \frac{p_{R1}-p_{R2}}{L} \int_L^{\infty} (s-L) s T(s) \D s.
\end{eqnarray}
To get from the third to the fourth equality, the order of the integrations over the region depicted in figure~\ref{figIntRegion}b was switched.

Finally, summing the two flux components, with $q_{R1,S} \equiv -q_{S,R1}$ determined by equations~\eq{eqqSR1} and~\eq{eqqSR12}, leads to
\begin{eqnarray}
q_{R1,S} + q_{R1,R2} & = & \frac{p_{R1}-p_{R2}}{L} \left[\int_0^L s^2 T(s) \D s + \int_L^\infty Ls T(s) \D s + \int_L^{\infty} (s-L) s T(s) \D s \right] \nonumber \\
 & = & \frac{p_{R1}-p_{R2}}{L} \int_0^\infty s^2 T(s) \D s = \frac{p_{R1}-p_{R2}}{L}\frac{k}{\mu},
\end{eqnarray}
where the last equality results from the Darcy limit~\eq{eqDarcyLimit}. Interestingly, given the assumptions made, the total flux from reservoir~$R_1$ through sample~$S$ to reservoir~$R_2$ is consistent with Darcy's law.

\subsection{Comparison with Pore-Network Results}\label{subsecNetworks}

\begin{figure}
\begin{center}
\includegraphics[width=0.55\textwidth]{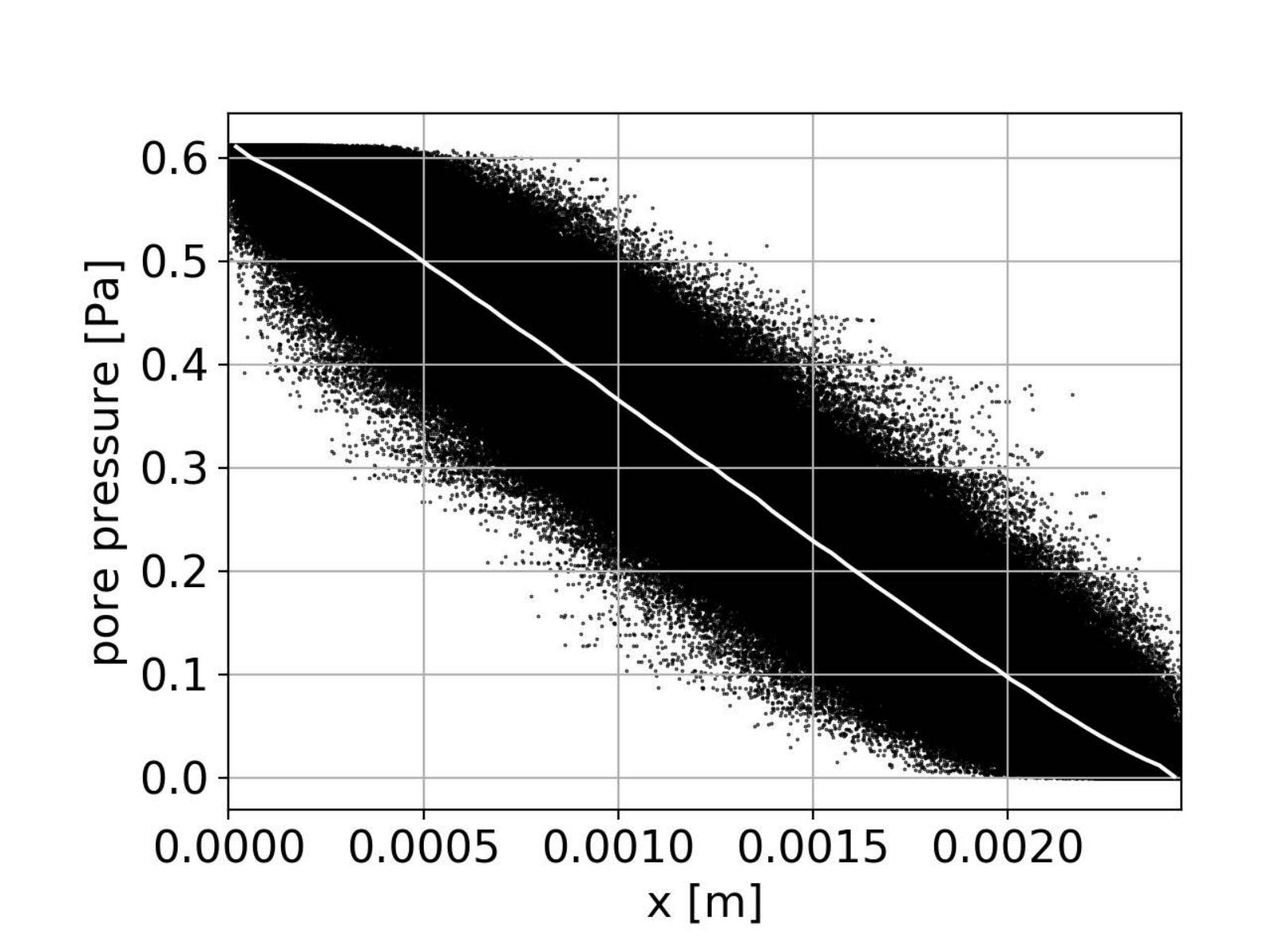}
\caption{Pore pressures in the sandstone network of thickness $L_x = 5L_m = 2.4\mbox{mm}$ with one million pores. Constant pressure boundary conditions were applied at throats at $x = 0$ and~$L_x$. The mean pressure curve resulting from averages over pores in 64 equidistant slabs perpendicular to the $x$-direction is depicted as a white line.\label{figPmPsBH1000}}
\end{center}
\end{figure}%
In a last part, the theoretical results derived in the previous sections shall be verified against numerical results from pore network calculations. In line with the sandstone network considered in section~\ref{subsecT}, we start our verification study with the same network type. By using the algorithm outlined in section~\ref{subsecNetwork} and the sandstone dataset from \citet{Idowu:2009b}, a periodic network of extensions $5L_m\times 59L_m\times 59L_m$ with $L_m = 0.489\mbox{mm}$, one million pores, and 2.1 million throats was generated (corresponding to an average coordination number of 4.2). In a next step, periodic throats in $x$-direction were cut open at the network boundaries at $x = 0$ and $L_x$ and constant pressures $p_{R1}$ and $p_{R2}$, respectively, were applied at the throat cutting points as sketched in figure~\ref{figBoundedNetwork}. The resulting pore pressures along the network are reported in figure~\ref{figPmPsBH1000}. While the mean pressure (white line) follows a linear trend in good agreement with the theoretical result reported in section~\ref{subsecPm}, local pore pressures vary significantly and are mostly far from the mean pressure.

\subsubsection{Exact Short-Circuit Flux}

\begin{figure}
\unitlength\textwidth
\begin{picture}(1,0.37)
%\put(0,0.375){\line(1,0){1}} % top
\put(0.0,0.0){\includegraphics[width=0.5\textwidth]{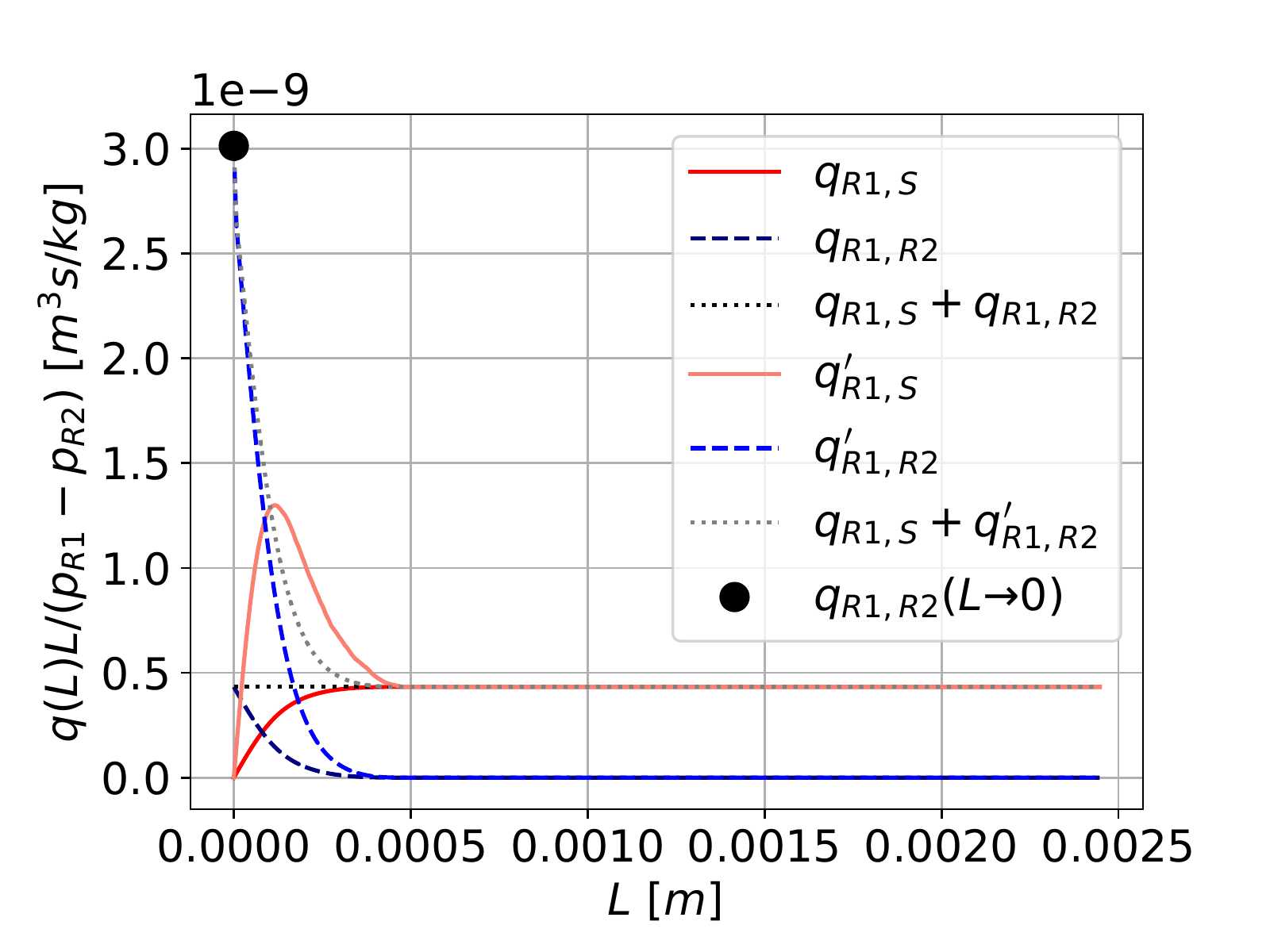}}
\put(0.47,0.36){\makebox(0,0){(a)}}
\put(0.5,0.0){\includegraphics[width=0.5\textwidth]{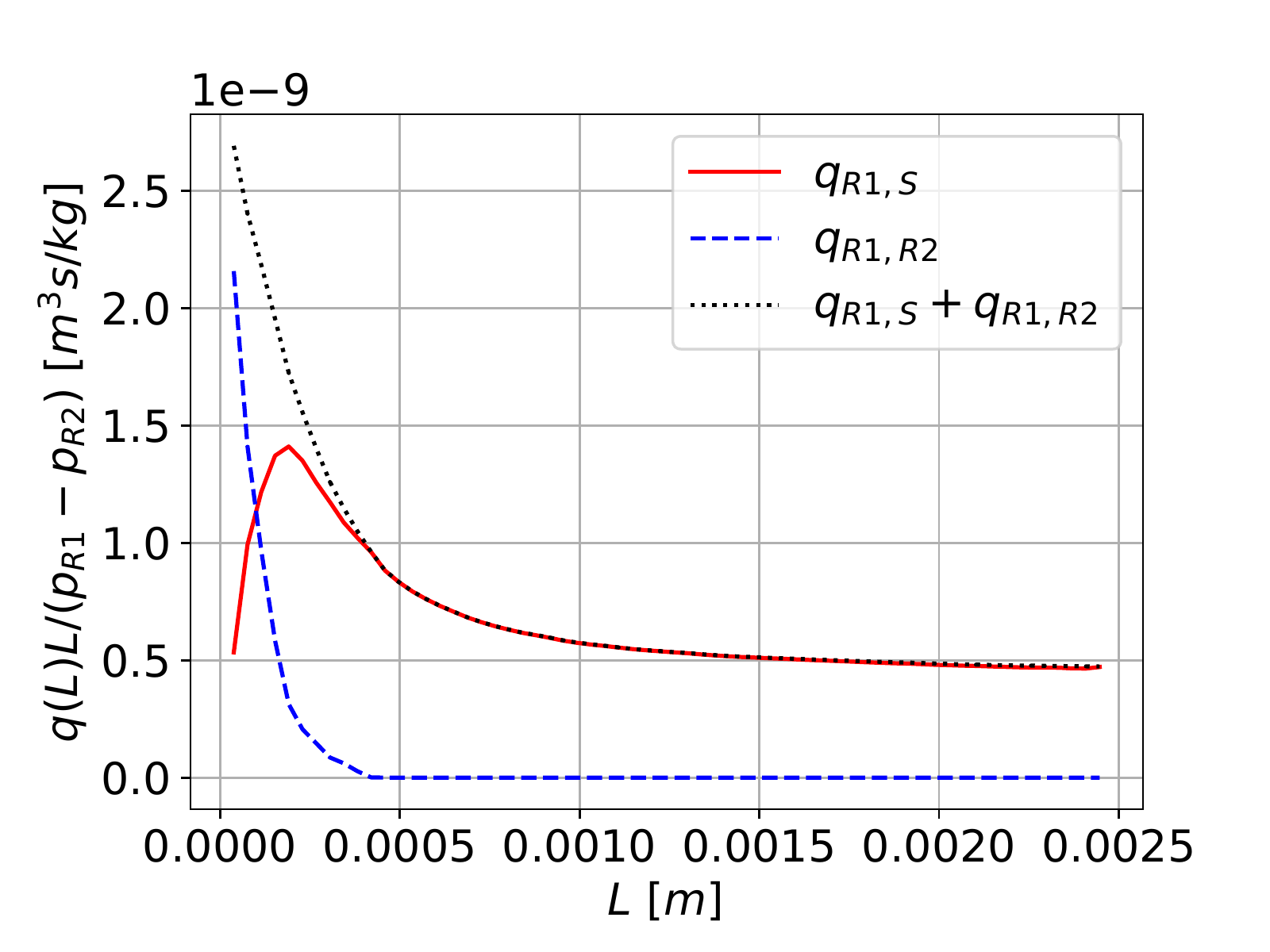}}
\put(0.97,0.36){\makebox(0,0){(b)}}
%\put(0,0.0){\line(1,0){1}} % bottom
\end{picture}
\caption{Normalized fluxes through sandstone networks of different thickness~$L$. For the largest network $L/L_m = 5$. Depicted are (a) theoretical results from sections~\ref{subsecQSR2} and~\ref{subsecQR1R2} as well as (b) numerical network results. $q_{R1,S}$ and $q_{R1,R2}$ represent the fluxes from reservoir~$R_1$ to pores in sample~$S$ (red solid) and from reservoir~$R_1$ through throats in sample~$S$ to reservoir~$R_2$ (blue dashed), respectively. The sum of the two fluxes (black dotted) is provided as well. $q_{R1,R2}(L\to 0)$ (black circle), $q^\prime_{R1,R2}$ (light blue dashed), and $q^\prime_{R1,S}$ (light red solid) given by equations~\eq{eqqR1R20}, \eq{eqqtR1R2}, and \eq{eqqtR1S}, respectively, are included in panel~(a).\label{figQBH1000}}
\end{figure}%
\begin{figure}
\unitlength\textwidth
\begin{picture}(1,0.37)
%\put(0,0.375){\line(1,0){1}} % top
\put(0.0,0.0){\includegraphics[width=0.5\textwidth]{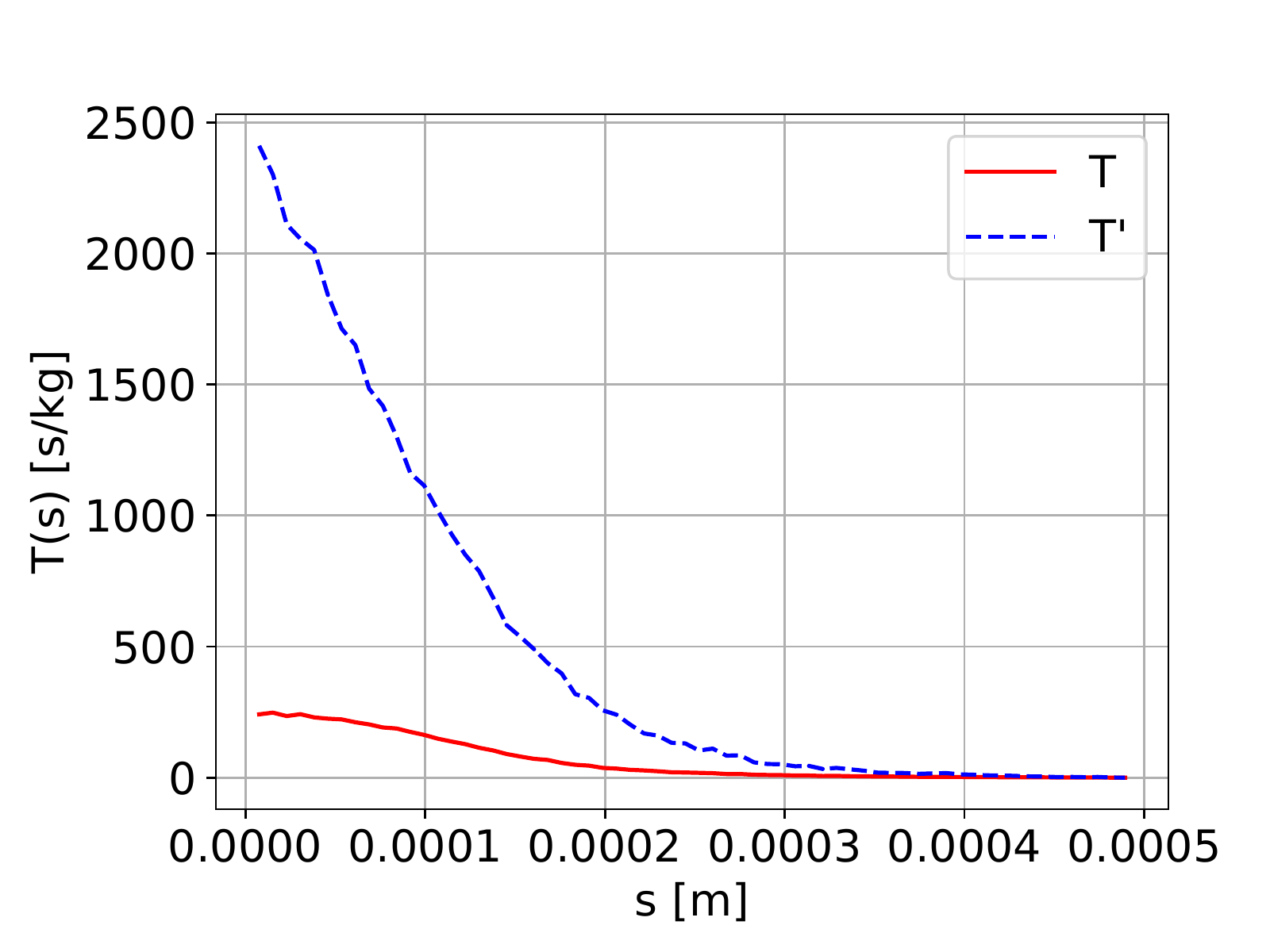}}
\put(0.47,0.36){\makebox(0,0){(a)}}
\put(0.5,0.0){\includegraphics[width=0.5\textwidth]{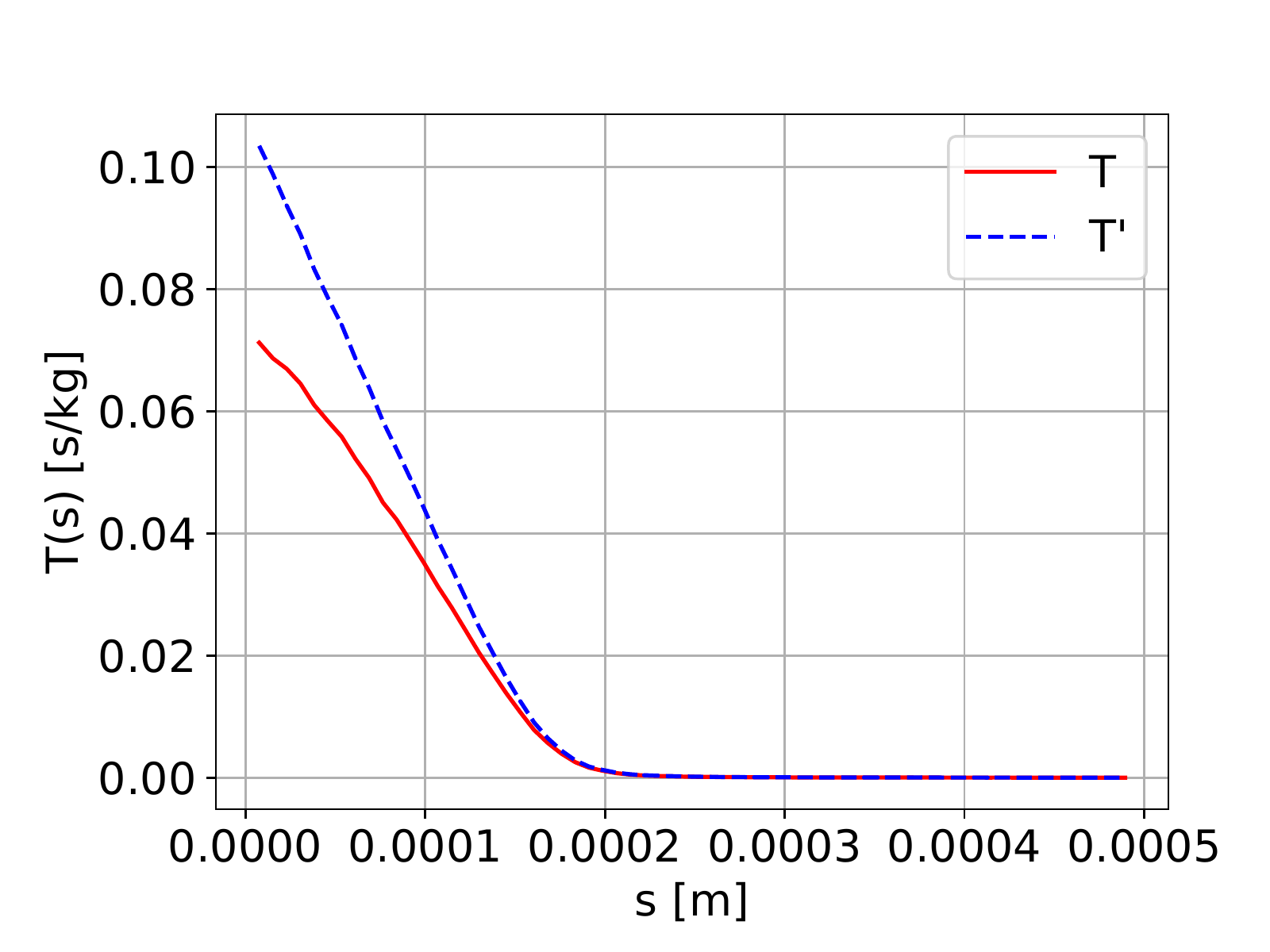}}
\put(0.97,0.36){\makebox(0,0){(b)}}
%\put(0,0.0){\line(1,0){1}} % bottom
\end{picture}
\caption{Conductivity distributions $T(s)$ (red solid) and $T^\prime(s)$ (blue dashed) as defined through expressions~\eq{eqTs} and \eq{eqTt}, respectively. In panels~(a) and (b), distributions resulting from the sandstone and the homogeneous network are depicted, respectively. In both networks, the longest throats have a length of $L_m = 4.89\times 10^{-4}\mbox{m}$.\label{figTBH1000C10}}
\end{figure}%
In order to assess the flux predictions from sections~\ref{subsecQSR2} and \ref{subsecQR1R2}, the pore network is gradually cut back leading to networks of decreasing thickness~$L_x$ or $L$. The resulting normalized fluxes are compared with their theoretical counterparts in figure~\ref{figQBH1000}. While the theoretical $q_{R1,S}$ correctly transitions from 0 to the asymptotic value $k/\mu = 3.86\times 10^{-13}\mbox{m}^2/[8.9\times 10^{-4}\mbox{kg/(ms)}] = 4.34\times 10^{-9}\mbox{m}^3 \mbox{s/kg}$ consistent with Darcy's law, it deviates from the numerical result at intermediate~$L$. Moreover, $q_{R1,R2}$ is considerably higher in networks with $L\to 0$ compared to the theoretical prediction. For a network with vanishing thickness~$L$, the likelihood to observe pores in the sample goes to zero (compare figure~\ref{figBoundedNetwork}). Consequently, the flux through the sample is simply given by
\begin{equation}\label{eqqR1R20}
q_{R1,R2}(L\to 0) = \frac{1}{C} \sum_{x_I < 0}\sum_{x_J > L} \frac{x_J - x_I}{L}T_{I,J}  (p_{R1}-p_{R2}) = \frac{p_{R1}-p_{R2}}{C L} \sum_{x_I < 0}\sum_{x_J > L} (x_J - x_I)T_{I,J}.
\end{equation}
Therefore as $L\to 0$, the limiting flux~\eq{eqqR1R20} becomes independent of the network thickness~$L$ for a constant pressure drop $(p_{R1}-p_{R2})/L$.\footnote{In expression~\eq{eqqR1R20}, $x_I$ and $x_J$ correspond to the pore positions of the uncut throats that go through sample~$S$.} In figure~\ref{figQBH1000}a, the limiting flux is included with a black circle being in agreement with the numerical limit seen in figure~\ref{figQBH1000}b. Motivated by the reasoning behind expression~\eq{eqqR1R20}, the approximate expression~\eq{eqqR1R2} for $q_{R1,R2}$ can be replaced by an exact relation: We write instead of equation~\eq{eqQD1D2}
\begin{equation}\label{eqQtD1D2}
Q^\prime_{D1,D2} = \sum_{I\in D_1} \sum_{J\in D_2} \frac{\overbrace{x_J-x_I}^{\displaystyle \hspace{-1em}\approx x_{D2}-x_{D1}\hspace{-1em}}}{L} T_{I,J}(\underbrace{p_I-p_J}_{\displaystyle \hspace{-1em}=p_{R1}-p_{R2}\hspace{-1em}}) = C \frac{x_{D2}-x_{D1}}{L} T^\prime(x_{D1}-x_{D2}) (p_{R1}-p_{R2}) h^2
\end{equation}
with the geometric conductivity distribution defined as
\begin{equation}\label{eqTt}
T^\prime(x_A-x_B) \equiv \frac{1}{Ch^2} \sum_{I\in A}\sum_{J\in B} T_{I,J}.
\end{equation}
$T^\prime(s)$ is determined by the network geometry, but in contrast to $T(s)$ it does not depend on the flow. A comparison of the flow- vs.\ the geometry-based conductivity distribution for the sandstone network is provided in figure~\ref{figTBH1000C10}a. Insertion of expression~\eq{eqQtD1D2} in derivation~\eq{eqqR1R2} in place of $Q_{D1,D2}$ leads to
\begin{equation}\label{eqqtR1R2}
q^\prime_{R1,R2} = \frac{p_{R1}-p_{R2}}{L} \int_L^{\infty} (s-L) s T^\prime(s) \D s.
\end{equation}
As seen in figure~\ref{figQBH1000}, this flux is in agreement with its numerical counterpart extracted from the sandstone network. Moreover, it is consistent with the flux limit~\eq{eqqR1R20}.

\subsubsection{Pressure and Pore Connectivitiy}

The remaining deviations in $q_{R1,S}$ are analyzed next by inspecting the simplifying assumption of setting pore pressures equal to mean slab pressures (compare derivation~\eq{eqQA}). In figure~\ref{figPmPsBH1000}, it was found for a sandstone network of thickness $L/L_m = 5$ that pore pressures deviate significantly from the local mean pressure. A similar analysis as with the sandstone network was performed with a beadpack network \citep{Bijeljic:2016a} displaying a higher pore connectivity than the sandstone. From the network data obtained from \citet{Bijeljic:2016a}, a beakpack network of extensions $4L_m\times 55L_m\times 55L_m$ with $L_m = 1.96\times 10^{-4}\mbox{m}$, $3\times 10^5$ pores, and one million throats was generated by following the steps listed at the beginning of section~\ref{subsecNetworks}. Moreover, a homogeneous network, where each pore is connected to ten neighboring pores through throats of equal radii, was studied. This network has extensions of $2L_m\times 41L_m\times 41L_m$ with $L_m = 4.89\times 10^{-4}\mbox{m}$ and is comprised of $2\times 10^5$ pores and one million throats.

\begin{figure}
\begin{center}
\includegraphics[width=0.55\textwidth]{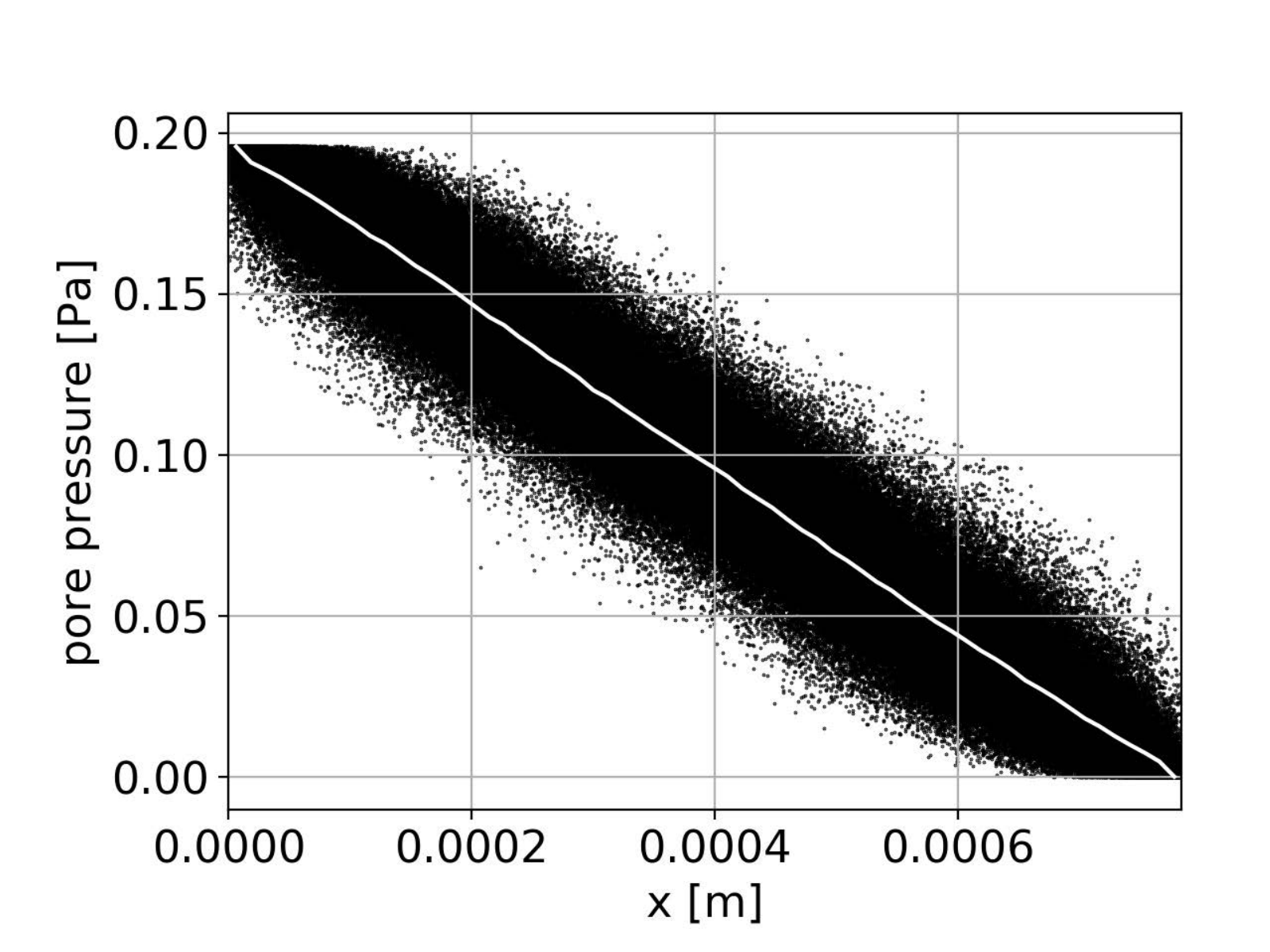}
\caption{Pore pressures in the beadpack network of thickness $L_x = 4L_m = 0.78\mbox{mm}$ with 0.3 million pores. See figure~\ref{figPmPsBH1000}.\label{figPmPsBP500}}
\end{center}
\end{figure}%
\begin{figure}
\unitlength\textwidth
\begin{picture}(1,0.37)
%\put(0,0.375){\line(1,0){1}} % top
\put(0.0,0.0){\includegraphics[width=0.5\textwidth]{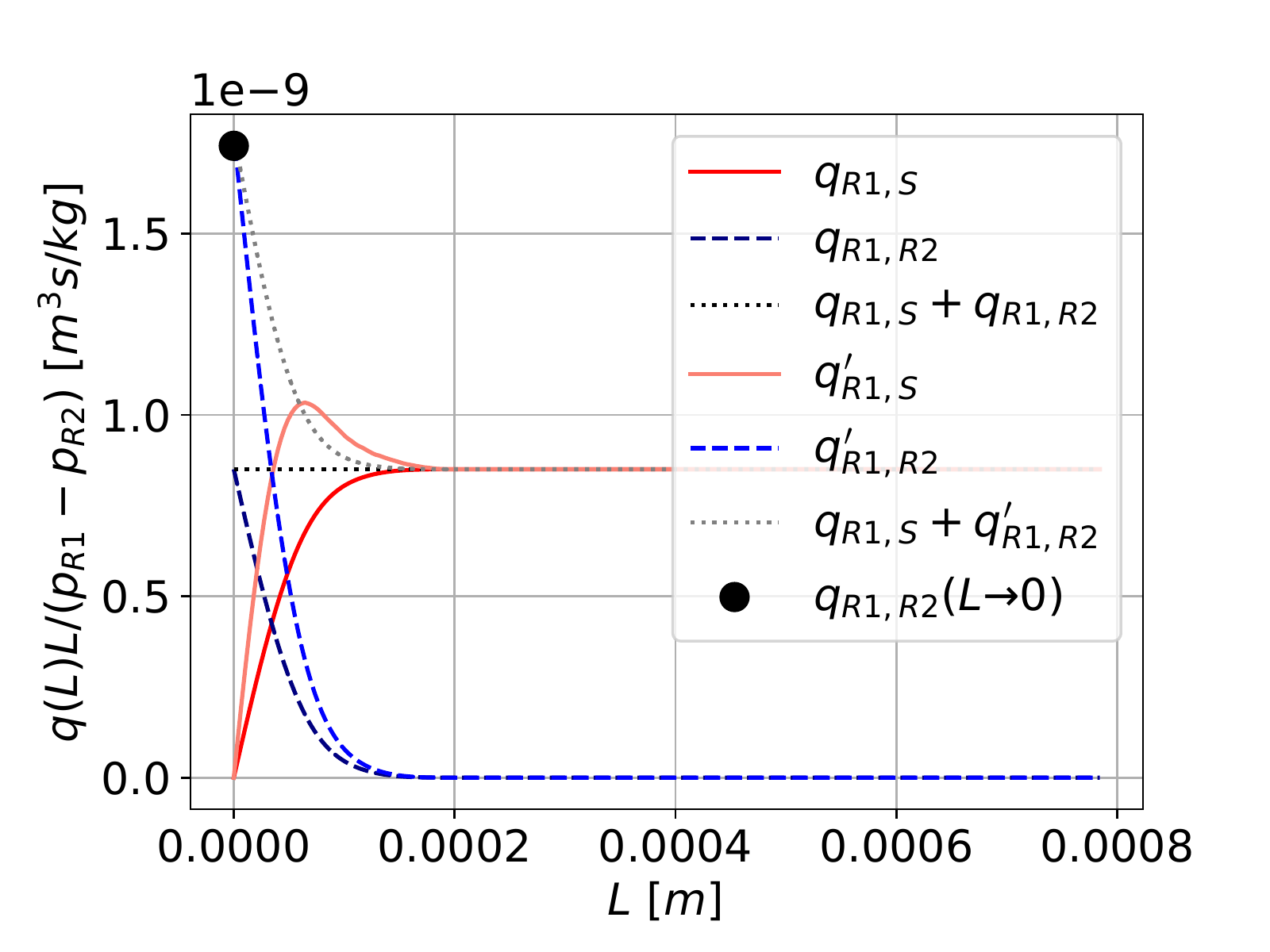}}
\put(0.47,0.36){\makebox(0,0){(a)}}
\put(0.5,0.0){\includegraphics[width=0.5\textwidth]{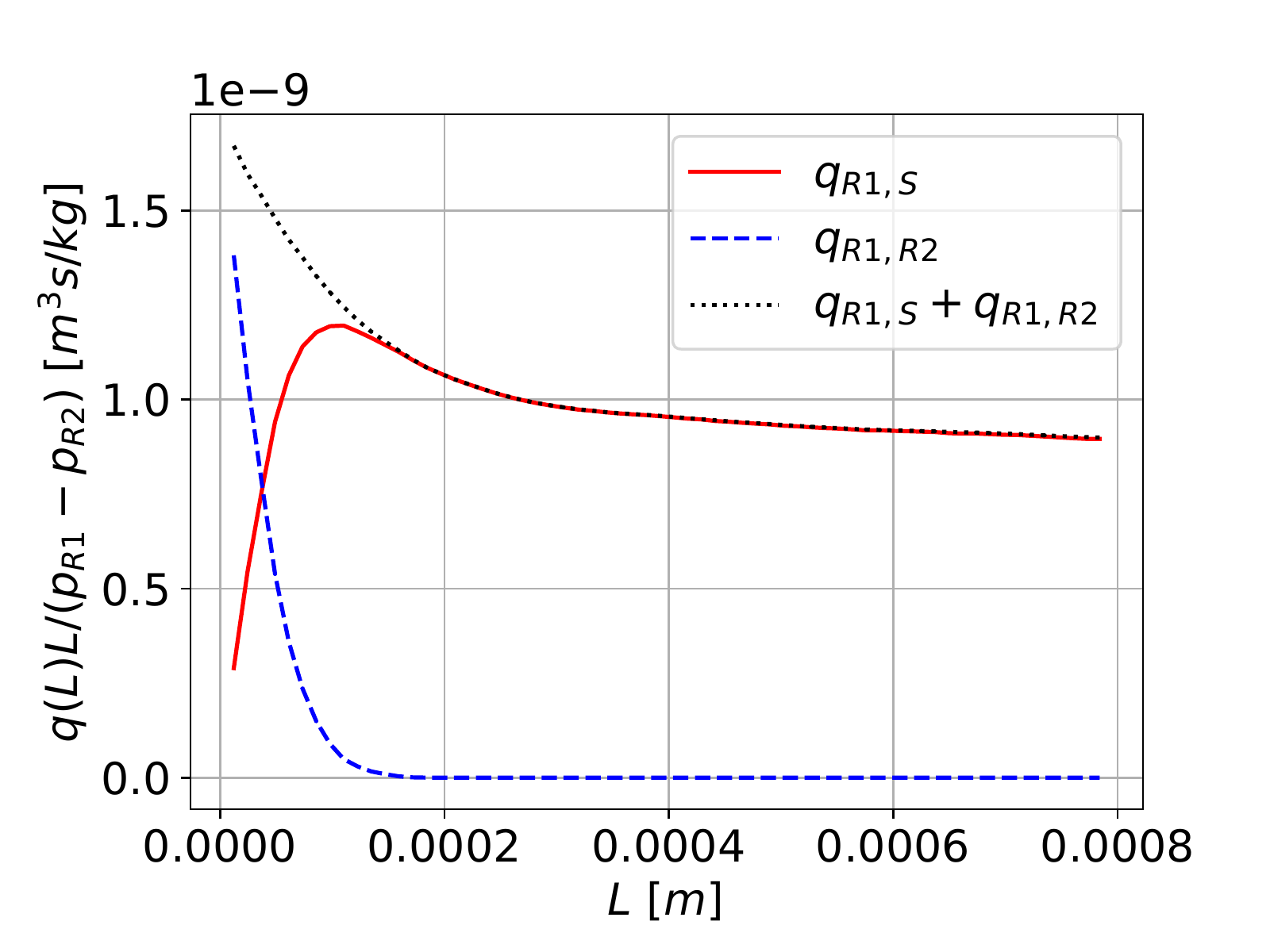}}
\put(0.97,0.36){\makebox(0,0){(b)}}
%\put(0,0.0){\line(1,0){1}} % bottom
\end{picture}
\caption{Normalized fluxes through beadpack networks of different thickness~$L$. For the largest network $L/L_m = 4$. For a data series description, see figure~\ref{figQBH1000}.\label{figQBP500}}
\end{figure}%
\begin{figure}
\begin{center}
\includegraphics[width=0.55\textwidth]{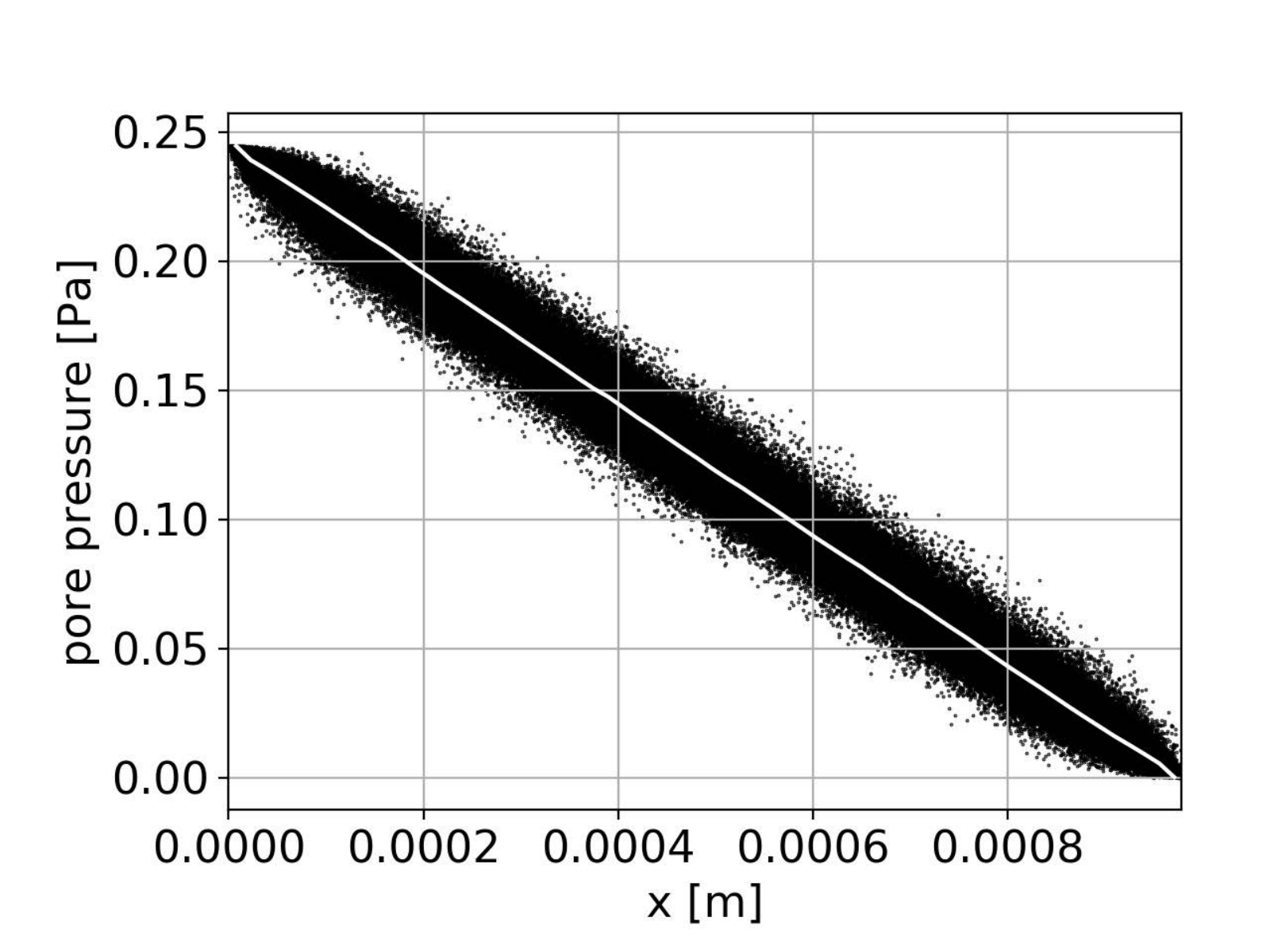}
\caption{Pore pressures in a homogeneous network with pore-coordination number 10 and thickness $L_x = 2L_m = 0.98\mbox{mm}$ with 0.2 million pores. See figure~\ref{figPmPsBH1000}.\label{figPmPsC10}}
\end{center}
\end{figure}%
\begin{figure}
\unitlength\textwidth
\begin{picture}(1,0.37)
%\put(0,0.375){\line(1,0){1}} % top
\put(0.0,0.0){\includegraphics[width=0.5\textwidth]{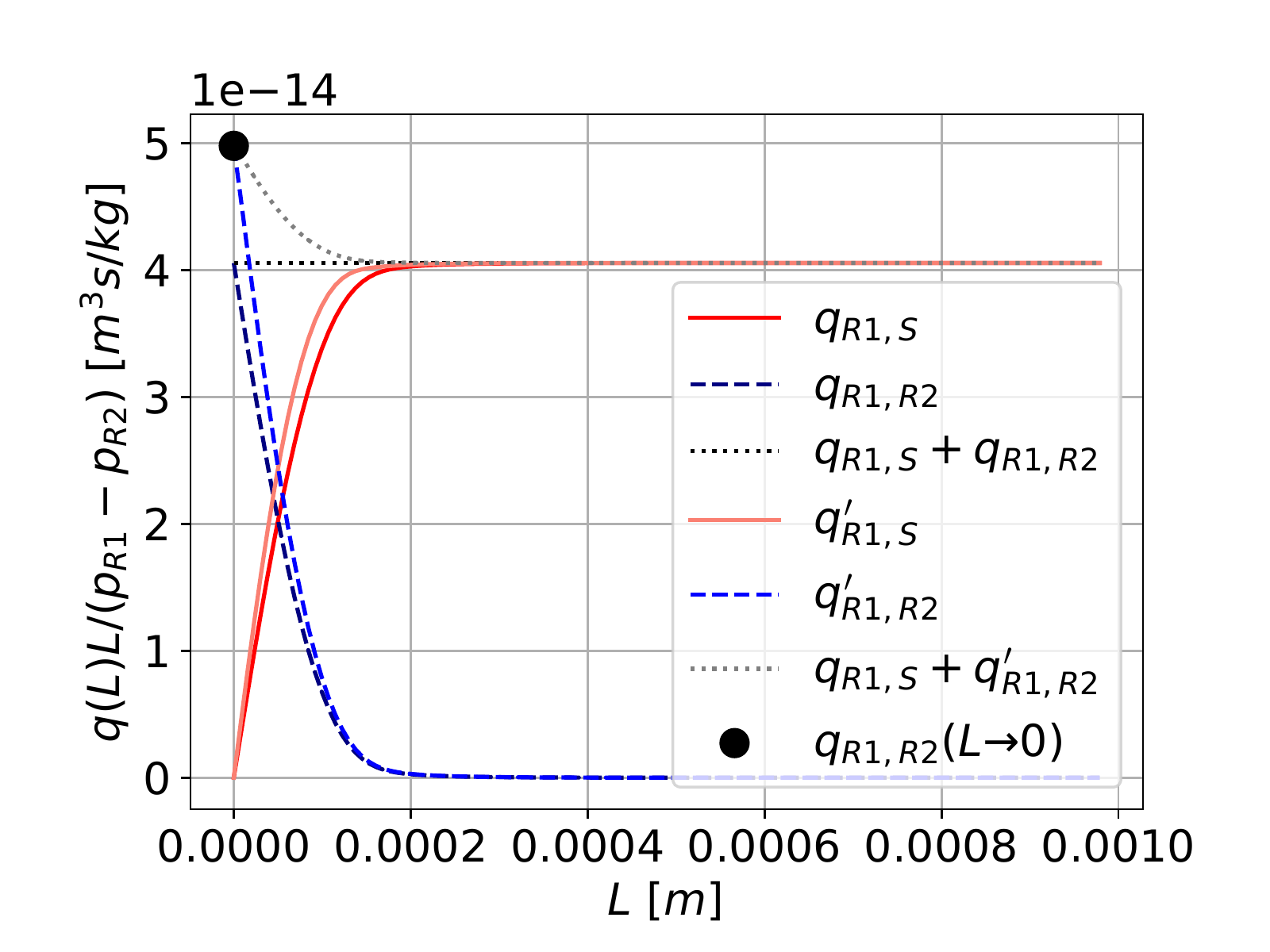}}
\put(0.47,0.36){\makebox(0,0){(a)}}
\put(0.5,0.0){\includegraphics[width=0.5\textwidth]{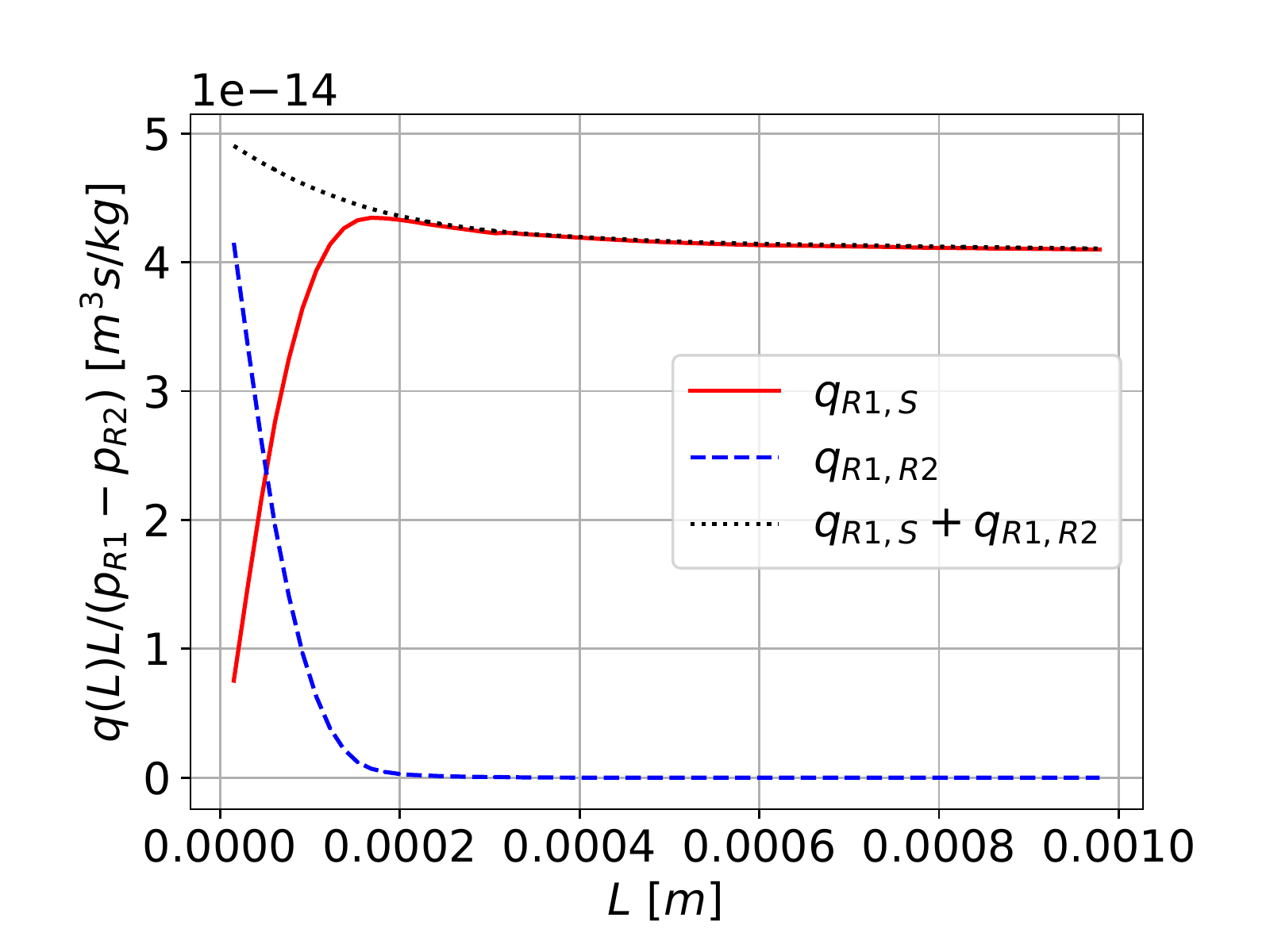}}
\put(0.97,0.36){\makebox(0,0){(b)}}
%\put(0,0.0){\line(1,0){1}} % bottom
\end{picture}
\caption{Normalized fluxes through homogeneous networks with pore-coordination number 10 and different thickness~$L$. For the largest network $L/L_m = 2$. For a data series description, see figure~\ref{figQBH1000}.\label{figQC10}}
\end{figure}%
The pore pressure variability for the beadpack and the homogeneous network are provided in figures~\ref{figPmPsBP500} and~\ref{figPmPsC10}, respectively. These results illustrate that an increased coordination number (average values of 6.6 for the beadpack and 10 for the homogeneous network) leads to more homogeneous pore pressures in better agreement with the assumptions made in our theoretical analysis. Moreover, the geometry- and flow-based conductivity distributions are more alike than for the sandstone network as partly seen in figure~\ref{figTBH1000C10}. Accordingly, the theoretical results are in better agreement with the corresponding pore network data as documented in figures~\ref{figQBP500} and~\ref{figQC10}. Even though the theoretical results in figure~\ref{figQC10}a come close to their numerical counterparts in panel b of that figure, $q_{R1,S}$ from the network does not only approach the Darcy limit from below, as predicted by the theory, but for $L \approx L_m$ exceeds the Darcy limit. This is related to the fact that for thin networks, pore pressures are closely correlated with the reservoir pressures (conductivity is quantified by $T^\prime(s)$) and thus vary locally less than in the space-stationary section emerging for large~$L$ in the center of the sample ($T(s)$ for conductivity). Thus, an ad-hoc improvement over the model~\eq{eqqSR1} for $q_{R1,S}$ (with $q_{R1,S} = -q_{S,R1}$) is given by
\begin{equation}\label{eqqtR1S}
q^\prime_{R1,S} \equiv \frac{p_{R1}-p_{R2}}{L} \left[ \int_0^L s^2 T(s) \D s + \int_L^\infty Ls T^\prime(s) \D s \right],
\end{equation}
which is included in figures~\ref{figQBH1000}a, \ref{figQBP500}a, and \ref{figQC10}a (light red solid lines). Here, in the second term $T(s)$ was replaced by $T^\prime(s)$. This term is large for small~$L$ and vanishes as $L\to L_m$, while the first term going to $k/\mu$ becomes dominant. Despite this modification, pore pressures remain correlated over several throat generations in flow direction~$x$ and thus the effect under inspection spans over lengths $L > L_m$, which is, however, not accounted for by $T^\prime(s)$ in expression~\eq{eqqtR1S}. Such long-range pressure effects are particularly pronounced in less-connected networks (see sandstone and beadpack results in figures~\ref{figQBH1000} and \ref{figQBP500}).

\section{Summary}

In the first part of this work, we have presented a method for the extraction of the flow-based conductivity distribution, which is at the heart of the non-local Darcy formulation. The conductivity distribution accounts for non-local pressure transmission resulting from throats which connect distant pores. Unlike the in-/outflow slab method, which induces a mean flow in the network through a singular source/sink pair, the new approach based on a space-stationary network and flow field does not require buffer layers that absorb or dampen flow non-stationarities resulting from the localized source/sink. A convergence study for different slab\footnote{Slabs are pore network sub-volumes with flow and pressure statistics that are to a good approximation statistically homogeneous in space.} widths was successfully conducted and consistency with the classical Darcy limit has been verified.

By assuming a stationary conductivity distribution and by approximating pore pressures based on their local slab means, a theory describing the flow in bounded networks has been derived. The theory implies a constant mean pressure gradient in the sample and a net flux through the sample both consistent with Darcy's law. In a next step, these theoretical results were verified with numerical network simulations. It was found that the stated assumptions are reasonable for highly-connected networks, but inadequate with realistic natural pore networks leading to inaccurate predictions. While approximately linear mean pore-pressure drops in agreement with the theory were found, high local pore-pressure variances and deviations in the total flux through thin networks were observed. In natural networks, the stated pressure variations are of similar size as the mean pressure drop over a characteristic network length (e.g., length of the longest throat~$L_m$). Since the pressure is fixed to a constant value at boundaries (inducing zero variance), the flow statistics near the bounds deviate from the interior of the network, where the flow is similar to a periodic network (with non-zero variance). Introduction of a second conductivity distribution based entirely on the network geometry, and thus accounting for the zero pressure variability at boundaries, leads to reasonably accurate theoretical flux predictions. In conclusion, the presented theory is expected to provide accurate results for networks where the conductivity distribution based on flow is similar to the one based on the network geometry, with the latter ignoring the coupling between pore pressures and throat conductivities.

One remaining open aspect of this work is the organization of the flow along sequentially linked throats that connect the interior of the network with its boundaries. This effect is not captured by the present theory based on the two listed conductivity distributions, but clearly observable in the numerical network results: While in the theory, changes in the fluxes were predicted up to networks of maximal thickness given by the longest throat~$L_m$ (both conductivity distributions have a spatial support of $2L_m$), in the numerical pore network results, flux dependencies for much thicker networks were found. It seems necessary to define an effective throat length that is longer than actual throats connecting just two pores. A simple example of the occurrence of an effective throat is implied by a throat pair that is linked through a common pore, which in turn has no further throat connections. Here, the effective throat results from the sequential combination of the throat pair, while the intermediate pore can be safely ignored.

\bibliographystyle{plainnat}

\bibliography{dwm_ssf}
\end{document}